\documentclass[manuscript]{acmart}
\AtBeginDocument{%
  \providecommand\BibTeX{{%
    \normalfont B\kfern-0.5em{\scshape i\kern-0.25em b}\kern-0.8em\TeX}}}
    
\setcopyright{acmcopyright}
\copyrightyear{2025}
\acmYear{2025}
\acmDOI{}
\acmJournal{TOSEM}
\acmMonth{12}

\usepackage{graphicx}
\usepackage{xurl}
\usepackage{booktabs}
\usepackage{multibib}
\usepackage{natbib}
\usepackage{hhline}
\usepackage{xcolor}
\usepackage{graphicx}
\usepackage{colortbl}
\usepackage{multirow}
\usepackage{tabularx}
\usepackage{siunitx}
\usepackage{pdflscape}
\usepackage{tabularx}
\usepackage{array}
\usepackage{ragged2e}
\usepackage{subcaption}

\begin{document}

\title{Taking a Pulse on How Generative AI is Reshaping the Software Engineering Research Landscape}

\author[B. Trinkenreich]{Bianca Trinkenreich}
  \email{bianca.trinkenreich@colostate.edu}
  \affiliation{%
  \institution{Colorado State University}
  \city{Fort Collins, CO}
  \country{USA}}

\author[F. Calefato]{Fabio Calefato}
\email{fabio.calefato@uniba.it}
\affiliation{%
\institution{University of Bari}
\city{Bari}
\country{Italy}}

\author[K. Blincoe]{Kelly Blincoe}
\email{k.blincoe@auckland.ac.nz}
\affiliation{%
\institution{University of Auckland}
\city{Auckland}
\country{New Zealand}}

\author[V.T. Wivestad]{Viggo Tellefsen Wivestad}
\email{viggo.wivestad@sintef.no}
\affiliation{%
  \institution{SINTEF Digital}
  \city{Trondheim}
  \country{Norway}}

\author[A.P. Santos Alves]{Antonio Pedro Santos Alves}
 \email{apsalves@inf.puc-rio.br}
\affiliation{%
  \institution{PUC-Rio}
  \city{Rio de Janeiro}
  \country{Brazil}}

\author[J. Condé Araújo]{Júlia Condé Araújo}
\email{julia.condearaujo@colostate.edu}
\affiliation{%
\institution{Colorado State University}
\city{Fort Collins}
\state{CO}
\country{USA}}

 \author[M. Condé Araújo]{Marina Condé Araújo}
 \email{marina.condearaujo@colostate.edu}
 \affiliation{%
   \institution{Colorado State University}
\city{Fort Collins}
\state{CO}
   \country{USA}}

\author[P. Tell]{Paolo Tell}
\email{pate@itu.dk}
\affiliation{%
  \institution{IT University of Copenhagen}
  \city{Copenhagen}
  \country{Denmark}
}

\author[M. Kalinowski]{Marcos Kalinowski}
\email{kalinowski@inf.puc-rio.br}
\affiliation{%
  \institution{PUC-Rio}
  \city{Rio de Janeiro}
  \country{Brazil}
}

\author[T. Zimmermann]{Thomas Zimmermann}
\email{tzimmer@uci.edu}
\affiliation{%
  \institution{University of California Irvine}
  \city{Irvine, CA}
  \country{USA}
}

\author[M-A. Storey]{Margaret-Anne Storey}
\email{mstorey@uvic.ca}
\affiliation{%
  \institution{University of Victoria}
  \city{Victoria}
  \country{Canada}
}

\begin{abstract}
\textbf{Context:} Software engineering (SE) researchers increasingly study Generative AI (GenAI) while also incorporating it into their own research practices. Despite rapid adoption, there is limited empirical evidence on how GenAI is used in SE research and its implications for research practices and governance. \textbf{Aims:} We conduct a large-scale survey of 457 SE researchers publishing in top venues (2023–2025). \textbf{Method:} Using quantitative and qualitative analyses, we examine who uses GenAI and why, where it is used across research activities, and how researchers perceive its benefits, opportunities, challenges, risks, and governance. \textbf{Results:} GenAI use is widespread, with many researchers reporting pressure to adopt and align their work with it. Usage is concentrated in writing and early-stage activities, while methodological and analytical tasks remain largely human-driven. Although productivity gains are widely perceived, concerns about trust, correctness, and regulatory uncertainty persist. Researchers highlight risks such as inaccuracies and bias, emphasize mitigation through human oversight and verification, and call for clearer governance, including guidance on responsible use and peer review. \textbf{Conclusion:} We provide a fine-grained, SE-specific characterization of GenAI use across research activities, along with taxonomies of GenAI use cases for research and peer review, opportunities, risks, mitigation strategies, and governance needs. These findings establish an empirical baseline for the responsible integration of GenAI into academic practice.

\end{abstract}

\maketitle
\renewcommand{\shortauthors}{Trinkenreich et al.}

\section{Introduction}

Generative AI (GenAI) is disrupting not only how we develop and engineer software~\cite{hou2024llm4se}, but also how we conduct research~\cite{xu2024aiforss, luo2025llm4sr}.  Increasingly, researchers are using GenAI tools to support a wide array of research practices, including literature review~\cite{khraisha2024can, syriani2024screening, huotala2024promise, felizardo2024chatgpt}, coding and data analysis~\cite{bano2024large, barros2025large, ornelas2025llm, lecca2025applications, rasheed2024llmdataanalysts}, writing and editing manuscripts~\cite{kobak2025llmwriting, mishra2024llmacademic}, and reviewing and summarizing research~\cite{zhou2024llmreviewer, liang2024llmfeedback}. GenAI is also reported to assist with research idea generation~\cite{si2025llmideageneration} and with learning how to conduct and evaluate research~\cite{luo2025llm4sr}.

The adoption of GenAI by researchers mirrors the rapid uptake of GenAI by software engineers~\cite{dora2025ai,github2025octoverse,stackoverflow2025survey}. As many software engineering (SE) researchers shift their research agendas to studying the impact of GenAI on SE, they are simultaneously becoming users of the technology they are studying~\cite{russo2024manifesto}.  This dual role raises important, and often controversial questions about how these tools should influence research practices and research outcomes, in particular regarding the reliability, transparency, and integrity of scholarly work. 


Despite growing adoption, there is still limited empirical evidence about how researchers are actually using these tools and how their use may influence research practices and outcomes in software engineering \cite{wivestad2025attitudes}. Despite some early guidelines~\cite{wagner2025towards,baltes2025guidelines}, many of our colleagues are unsure what will be acceptable in using this technology.  Developing early evidence is essential for understanding both the opportunities GenAI offers and the risks it may introduce, and for informing responsible use as well as emerging policies within academic institutions and publishing organizations. Research practices often evolve more slowly than comparable practices in industry~\cite{shaw2002makes}, and yet researchers also face some of the same productivity pressures other software development professionals feel to adopt and use GenAI~\cite{miller2026}. 

To better understand these issues and mounting concerns, we conducted a large-scale survey of software engineering researchers.
Through a rigorously designed online survey instrument, we aimed to answer the following research questions: 
\begin{itemize}
    \item RQ1: Who is using GenAI in SE research and what motivates their use?
    \item RQ2: Where and how are researchers in SE using GenAI? 
    \item RQ3: What benefits, challenges, and opportunities do SE researchers perceive when using GenAI?
    \item RQ4: How do SE researchers trust its use, what risks do they perceive from using GenAI in research, and how do they mitigate those risks? 
    \item RQ5: What regulations and policies do SE researchers feel should govern the use of GenAI in research? 
\end{itemize}

Our survey instrument builds on prior survey research and theories, while it also probes into specifics about how GenAI is used and how its use is perceived in SE research.  The survey includes a mix of closed and open questions and was distributed to the authors of research papers published in the top venues in software engineering between 2023 to 2025.  We received 457 responses from researchers with varying experience levels from institutions around the world. 

Using a combination of quantitative analysis and qualitative inductive and deductive analysis, we make the following contributions: 
\begin{itemize}
    \item We provide one of the first empirical characterizations of how GenAI is being used in software engineering research and which research activities are seeing the most early adoption. 
    \item 
    We develop a taxonomy of the benefits, challenges, and opportunities SE researchers perceive from using GenAI. 
    \item 
    We identify and surface key risks SE researchers are concerned about, and examine how researchers build trust with using GenAI in SE research. We synthesize how they mitigate those risks, such as maintaining the human in the loop, set boundaries on tool usage, and aim to improve GenAI education.
    \item Finally, we report SE researchers' perspectives on what regulations and policies they believe should be in place for understanding the implications of using GenAI in research and peer review.
\end{itemize}

Our findings uncover critical tensions surrounding the use of GenAI in research and highlight important implications for research conducted in both academia and industry using GenAI. As GenAI models and tools continue to evolve, and as community expectations and social norms mature, longitudinal studies will be needed to understand how researchers’ practices and perceptions change over time. Our work provides a baseline for such future investigations and may help shape and guide how GenAI is used in the SE research community.

    



\section{Background and Related Work}

In the past few years, there are several papers that explore the role of Generative AI (GenAI) and large language models (LLMs) in research, and many that specifically consider research in software engineering (SE). 
%
%
We organize this prior work in four clusters: surveys that investigate the adoption and perceptions of GenAI use in research; papers that discuss methodological implications and propose guidelines for GenAI use in research; the use of LLMs to support specific research tasks;  and studies that explore the use of LLMs as research subjects. We begin with survey-based evidence on adoption and perceptions, which directly motivates our study.

\paragraph{Adoption and Perceptions of GenAI for Research}

The idea of using AI to augment scientific discovery predates modern generative models. Early work argued that the increasing complexity of scientific workflows creates bottlenecks that AI systems could help alleviate, particularly in tasks such as hypothesis generation and literature analysis~\cite{gil2014amplify}. Recent large-scale surveys provide initial empirical evidence of how researchers are adopting GenAI in practice. A global survey conducted by Nature reported that researchers (across many scientific domains) are experimenting with GenAI for idea generation, coding, and manuscript preparation~\cite{van2023ai}, while also raising concerns about misinformation, plagiarism, and inaccuracies in research outputs~\cite{van2023ai}. In a survey of Danish researchers from various domains, Andersen et al.~\cite{andersen2025generative} examined GenAI use across stages of the research process and found that researchers perceive clear benefits for writing-related tasks but express reservations when GenAI is applied to activities requiring methodological rigor, such as experimental design. Focusing specifically on the use of GenAI in SE research at a large European research institute, Wivestad and Barbala~\cite{wivestad2025attitudes} showed that LLM use is generally considered acceptable for narrow, verifiable tasks, but becomes more controversial in high-stakes contexts such as peer review. The studies also reported different adoption and perceptions across experience levels, with early-career researchers more likely to adopt GenAI~\cite{andersen2025generative}, while more experienced researchers emphasize risks related to rigor and reproducibility~\cite{wivestad2025attitudes}.

\paragraph{Methodological Reflections and Evaluation Guidelines for SE Researchers}

As GenAI becomes part of SE research workflows, questions about how to use these tools rigorously have gained attention. Prior work emphasized the need to report model versions, prompts, and configurations to enable reproducibility and comparability~\cite{baltes2025guidelines,wagner2025towards}. The non-deterministic and opaque nature of LLMs makes replication hard when such details are not documented~\cite{wagner2025towards}. Prior work also highlighted that introducing LLMs in research may destabilize core research constructs, as notions such as “developer”, “artifact”, and “interaction” become blurred in settings where humans and AI systems co-create software artifacts~\cite{treude2025generative}. This shift complicates the attribution of agency and what is being observed and measured. LLMs also introduce new data modalities, such as prompts and AI-generated artifacts, which raise concerns about bias, provenance, and interpretability. Furthermore, the variability of LLM outputs and the rapid evolution of models introduce evaluation drift, weakening reproducibility and causal inference~\cite{treude2025generative}. Finally, the use of LLMs as research instruments raises risks of over-reliance on automated analysis and potential loss of critical human judgment. Trinkenreich et al.~\cite{trinkenreich2025get} and Williams et al.~\cite{williams2025empirical} further argue that efficiency gains from GenAI must be balanced with safeguards to preserve rigor and transparency.

\paragraph{LLMs as Research Assistants in Empirical SE}

Beyond methodological considerations, several studies examined how LLMs are used to support specific SE research tasks. In qualitative analysis, LLMs have been used to support coding and theme generation at scale~\cite{montes2025large,ornelas2025llm}. Results show that their outputs can align with human-generated codes in clarity and organization, but they may also be overly granular and fragmented when models fail to capture latent meaning and produce coherent higher-level themes~\cite{montes2025large}. Similarly, LLM-assisted analysis can lead to loss of contextual depth and premature closure of interpretation, as models favor surface-level patterns over nuanced reasoning~\cite{ornelas2025llm}. Across both studies, LLM outputs are highly sensitive to prompt design and require iterative prompting and human validation, reinforcing their role as assistive rather than authoritative tools in qualitative analysis~\cite{ornelas2025llm,montes2025large}. In secondary studies, such as systematic literature reviews, LLMs have been applied to tasks including abstract screening, study selection, and data extraction, reducing manual effort and accelerating the processing of large corpora~\cite{huotala2024promise,felizardo2024chatgpt,felizardo2025difficulties}. Empirical evaluations show that LLMs can achieve moderate to high accuracy in study selection, but still produce incorrect classifications, including false negatives that may lead to loss of relevant evidence~\cite{felizardo2024chatgpt}. At the same time, LLM-assisted screening has been shown as not consistently more accurate than human reviewers, with performance depending on model choice and prompting strategies~\cite{huotala2024promise}. Prior work also highlighted methodological challenges, including sensitivity to prompt design, limited contextual information, and lack of transparency and reproducibility due to model variability and configuration changes~\cite{felizardo2025difficulties}. Across these studies, LLMs reduce effort but do not replace human judgment, requiring oversight to ensure accuracy and completeness of the review process~\cite{huotala2024promise,felizardo2024chatgpt}.

LLMs have also been explored for annotation and text processing, where model-generated annotations can be effective under controlled conditions~\cite{ahmed2025can}, particularly when supported by validation procedures~\cite{imran2025olaf}. For sentiment analysis of software engineering data, large models achieve competitive performance in some scenarios, although traditional fine-tuned models may still outperform them in others~\cite{zhang2025revisiting}.

\paragraph{LLMs as Research Subjects for SE Studies}

Other studies focus on LLMs themselves, examining whether they can simulate research processes or replace human participants. Liang et al.~\cite{liang2024can} show that LLMs can reproduce aspects of research workflows, but struggle with implicit reasoning and methodological nuance. Research on synthetic participants indicated that LLM-generated responses may be useful for exploratory purposes, but differ from human data in variability, depth, and contextual grounding~\cite{steinmacher2024can,gerosa2024can}. Harding et al.~\cite{harding2024ai} and De et al.~\cite{de2025investigation} warned that treating synthetic outputs as equivalent to human data can distort findings by reducing variability and inflating agreement.

\paragraph{Research Gap and Study Motivation}

Prior work has brought early insights on how GenAI is being used by researchers (state of the practice) as well as how GenAI should be used and can be used by researchers (state of the art). 

Our study is directly inspired by prior surveys that explored the adoption and perceptions of researchers of GenAI~\cite{van2023ai,andersen2025generative,wivestad2025attitudes}. We build on these initial findings and provide an SE-specific characterization and a detailed pulse of how GenAI is being used by a broad population of SE researchers around the world. We examine not only adoption and perceptions, but also how GenAI is used across SE research strategies, stages, and activities; the benefits, challenges, opportunities, and risks researchers associate with its use; how SE researchers establish trust and mitigate risks; and how they view regulation and the use of GenAI in SE research peer review. 
The next section describes the survey design, recruitment strategy, and analysis procedures.

\section{Research Method}
We describe the method followed by presenting the research questions in Section \ref{sec:method:questions}, the instrument design and data collection procedures in Section \ref{sec:method:design}, and the detailed description of the data analysis procedures in Section \ref{sec:method:analysis}.  Information regarding the replication package are provided in Section \ref{sec:method:replication}.

\subsection{Research objective and research questions}\label{sec:method:questions}
The objective of this research is to understand how GenAI is used in software engineering (SE) research. Specifically, this study aims to characterize \textit{who} among SE researchers is using GenAI, \textit{why} they are motivated to adopt it, \textit{when} and \textit{where} in the activities of the research pipeline they adopt it, and \textit{how} GenAI is incorporated into the research process. 
By providing an evidence-based characterization of current practices, this work seeks to inform ongoing methodological discussions on the role of GenAI in SE research and to establish a foundation for future normative and evaluative studies. 
For this, we address the following research questions.


\paragraph{\textbf{RQ1: Who is using GenAI in SE research and what is their motivation?}} This research question aims to identify who adopts GenAI providing context for interpreting patterns. To this end, \textbf{RQ1.1} investigates which SE researchers use GenAI, considering demographic and professional characteristics, such as research context, years of experience, and career stage.  The \textbf{RQ1.2} examines the motivations that drive researchers to use GenAI, including their perception of the current and anticipated impact of GenAI, as well as their perceived pressure on using and learning GenAI to stay relevant and to link their research to it. 

\paragraph {\textbf{RQ2: Where and how are researchers in SE using GenAI?}} This research question focuses on the concrete use of GenAI within the SE research process and peer review. 
Specifically, \textbf{RQ2.1} examines \textit{where}, i.e., in which research strategies, methods, and stages of the research pipeline, GenAI is being employed. \textbf{RQ2.2} provides a qualitative, fine-grained analysis of \textit{how} researchers use GenAI within these methods.  Finally, \textbf{RQ2.3} examines researchers' prior use and attitudes toward the use of GenAI in the peer-review process, a particularly sensitive area with implications for research integrity.

\paragraph {\textbf{RQ3:What benefits, challenges, and opportunities do SE researchers perceive when using GenAI?}}
\par \noindent
This research question investigates the perceived impact of GenAI on SE research, complementing the descriptive characterization of usage practices (RQ2) with researchers' assessments of its consequences. Understanding perceived impacts is essential for interpreting current adoption and for anticipating how GenAI may shape future research practices. In particular, \textbf{RQ3.1} examines the benefits that SE researchers perceive GenAI to offer. \textbf{RQ3.2} investigates the challenges researchers experience when using GenAI in their research. 
\textbf{RQ3.3} explores the opportunities researchers envision GenAI may offer for advancing SE research in the future. 

\paragraph {\textbf{RQ4: How do SE researchers trust its use and what risks do they perceive from using GenAI in research, and how do they mitigate those risks?}} While RQ3 focuses on perceived impacts, this research question explicitly addresses the risks associated with using GenAI in SE research and how researchers reason about managing these risks. Understanding risk perception and mitigation strategies is critical for informing responsible and sustainable research practices.
To this end, \textbf{RQ4.1} explores how researchers trust GenAI in different contexts and with regard to each research activity, \textbf{RQ4.2} examines how SE researchers perceive the risks of using GenAI for research, including concerns related to fabrication, plagiarism, quality, proliferation of misinformation, carbon footprint, and more. 
\textbf{RQ4.3} presents the strategies proposed by SE researchers to mitigate the risks associated with using GenAI. 
 

\paragraph {\textbf{RQ5: What regulations and policies do researchers feel should be applied to the use of GenAI in their research and reviewing activities?}} This research question focuses on normative and governance-related aspects of GenAI use in SE research. As GenAI adoption raises questions about appropriate boundaries and oversight, understanding researchers’ perspectives on regulation and policy is essential for informing institutional and community-level responses.
We investigate, through an open question, how researchers perceive the need to regulate the use of GenAI in research.


\subsection{Instrument Design}
\label{sec:method:design}
To collect empirical data for this study, we used a survey-based research design.  
We developed an online questionnaire~\footnote{The research protocol was approved by the Colorado State University institutional review board (IRB).} using Qualtrics\footnote{\url{http://www.qualtrics.com}} to collect detailed information from software engineering researchers regarding their use of generative AI technologies. The unit of analysis in this study is the individual researcher.
The set of questions comprising the questionnaire emerged from discussions during remote weekly meetings held by the authors of this paper over six months (December 2024 to June 2025).

\subsubsection{Instrument Structure}
After the consent form, the questionnaire included 17 questions related to our research questions, and five demographic questions for segmented analysis of the results. We used existing instruments where possible. The source for each question is presented in Table \ref{tab:rq-questionnaire}.

\newcommand{\quant}{\textit{[Quantitative]}}
\newcommand{\qual}{\textit{[Qualitative]}}

\begin{table}[t]
\caption{Research questions with corresponding questionnaire items, analysis method [\textit{Quantitative}|\textit{Qualitative}], and reference to the source for adapted items.}
\label{tab:rq-questionnaire}
\centering
\footnotesize
\begin{tabular}{p{.5cm}p{.9\textwidth}}
\toprule
    \multicolumn{2}{l}{\textbf{Research Questions}} \\
    & \textbf{Questionnaire Item. \textit{[Analysis]} } \\ 
\midrule
\multicolumn{2}{l}{\textbf{RQ1. Who is using GenAI in SE research and what is their motivation?}} \\
    \multicolumn{2}{l}{RQ1.1. Who (which researchers in SE) is using GenAI for their research?} \\
        & Demographics \quant \\
    \multicolumn{2}{l}{RQ1.2. What motivates researchers to use GenAI?} \\
        &How much of an impact do you think GenAI has or will have on SE Research? \quant \\
        &Do you feel under pressure to use and learn about GenAI for SE Research to stay relevant? \cite{wivestad2025attitudes} \quant \\
        &Do you feel under pressure to link your research to GenAI, or collaborate with AI researchers, in order to stay relevant, progress in your field, or secure funding? \cite{wivestad2025attitudes} \quant \\
\\[-7pt]
\multicolumn{2}{l}{\textbf{RQ2. Where and how are researchers in SE using GenAI?}} \\
    \multicolumn{2}{l}{RQ2.1. Where are researchers in SE using GenAI (research strategy, research method, and research pipeline stage)?} \\
        &More specifically, please indicate if you use GenAI for each of the following research activities (columns) across the different strategies (rows) \cite{trinkenreich2025get}. \quant \\
    \multicolumn{2}{l}{RQ2.2. How are researchers in SE using GenAI?} \\
        &Please briefly describe how you use GenAI in your research. \qual \\
     \multicolumn{2}{l}{RQ2.3. How do researchers feel about the use of GenAI in reviewing research papers?}\\
        &As a reviewer, have you used GenAI to assist in reviewing a SE research paper? \qual \\
        &Should reviewers be allowed or encouraged to use GenAI to assist in reviewing SE research papers? \qual \\ 
\\[-7pt]
\multicolumn{2}{l}{\textbf{RQ3. What benefits, challenges, and opportunities do SE researchers perceive when using GenAI?}} \\
    \multicolumn{2}{l}{RQ3.1. Which benefits do SE researchers perceive GenAI offers to SE Research?}\\
        &What benefits do you see GenAI bringing to SE Research? \cite{van2023ai} \quant \\
    \multicolumn{2}{l}{RQ3.2. What challenges do SE researchers experience using GenAI in SE research?} \\
        &What challenges (if any) do you, or your research team, face while using GenAI for SE Research \cite{wivestad2025attitudes}. \\
    \multicolumn{2}{l}{RQ3.3. What opportunities do SE researchers envision GenAI offers to SE Research?} \\
        &What opportunities do you see GenAI bringing to SE Research? \qual \\
\\[-7pt]
\multicolumn{2}{l}{\parbox{14.5cm}{\textbf{RQ4. How do SE researchers trust its use and what risks do they perceive from using GenAI in research, and how do they mitigate those risks? }}} \\
    \multicolumn{2}{l}{RQ4.1. How do SE researchers trust GenAI for research?} \\
    &For the following statements, please indicate your level of agreement on trust in using GenAI for SE research in general. \quant \\
    &For the following activities, I trust the use of GenAI \cite{choudhuri2025guides}. \\
    \multicolumn{2}{l}{RQ4.2. How do SE researchers perceive the risks of using GenAI for research?} \\
    &For the following statements, please indicate your level of agreement on the following risks that GenAI brings to SE Research? \cite{van2023ai} \quant \\
    \multicolumn{2}{l}{RQ4.3. How can SE researchers mitigate the risks while not missing the opportunities that GenAI offers?} \\
    &If you indicated being concerned about any risks in the previous question, how would you go about mitigating each of those risks? \qual \\

\\[-7pt]
\multicolumn{2}{l}{\textbf{RQ5. What regulations and policies do researchers feel should be applied to the use of GenAI in their research?}} \\
        &Should GenAI use be regulated in Software Engineering research, assuming that it is possible? (please elaborate) \qual \\
   
\\[-7pt]
\multicolumn{2}{l}{\textbf{Final thoughts}} \\ 
        &If you have final thoughts about using GenAI in SE research that might not have been covered in this questionnaire, please enter them here.  \qual \\
\bottomrule
\vspace{-5mm}
\end{tabular}

\end{table}


\subsubsection{Data Collection and Recruitment}
We employed a purposive sampling strategy~\cite{baltesESE2022-sampling} by recruiting authors of papers and articles published between 2023  and 2025 within a select set of leading software engineering venues: the International Conference on Software Engineering (ICSE), International Conference on Automated Software Engineering (ASE), International Conference on the Foundations of Software Engineering (FSE), IEEE Transactions on Software Engineering (TSE), ACM Transactions on Software Engineering and Methodology (TOSEM), and Springer Empirical Software Engineering (EMSE). 
We identified potential participants from the conference proceedings and the journal volumes, and contacted them via email with an invitation to participate in the survey. 
This strategy was chosen to ensure that respondents had demonstrable experience in software engineering research, aligning with the target population of the study.

The survey was initially distributed during the $1^{st}$ HumanAISE Workshop on Human-Centered AI for Software Engineering, held in Trondheim, Norway, on June 27, 2025. 
Then, three waves of email invitations were sent on July 20th, August 18th, and August 20th, 2025. Data collection was closed in September 2025, yielding 457 responses overall.

\subsection{Data analysis}
\label{sec:method:analysis}


\subsubsection{Filtering} 

\leavevmode\\
For each analysis, we applied item-level deletion by excluding responses with missing values for the respective question. No imputation was performed; hence, the reported sample size (\textit{n}) varies across analyses depending on the number of valid responses.

Unless otherwise specified, all plots and statistical summaries are based on the full set of valid responses to the respective question. The only exception is the analysis of GenAI use cases (Sec.~\ref{sec:rq2.2}), which was restricted to respondents who indicated they use GenAI for research purposes, filtering out researchers who stated using GenAI for other activities.

\subsubsection{Closed questions}


\paragraph{Overview}
We analyzed responses to closed-ended questions using descriptive statistics. Given the exploratory nature of this study and its goal of characterizing current practices and perceptions of GenAI use in SE research, we focused on summarizing the distribution of responses rather than conducting inferential statistical tests.

For categorical questions (e.g., use of GenAI across activities, perceived challenges), we report absolute counts and proportions of responses. For Likert-scale items (e.g., perceived benefits, risks, and trust), we computed the distribution of responses across all scale points and report percentages to facilitate comparison across groups.

To support interpretation, for Likert-scale items, we report Top-2-Box (e.g., \textit{Agree} + \textit{Completely Agree}) and Bottom 2 Box (e.g., \textit{Disagree} + \textit{Completely Disagree}) scores when appropriate, allowing us to summarize overall positive and negative perceptions.

All descriptive analyses and visualizations were generated using the set of valid responses for each question after applying the filtering procedures described in the previous section.

\paragraph{Segmented analysis}
Respondents were categorized based on their responses to which activities they use GenAI for. This was a multiple-choice question that listed several possible use cases, including both research- and non-research-related activities. Based on the response patterns, we operationalized three mutually exclusive subgroups:
\begin{enumerate}
\item Researchers who have used GenAI for research purposes (n=339)
\item Researchers who have used GenAI, but not for research (n=44)
\item Researchers who have not used GenAI (n=29)
\end{enumerate}

This grouping enabled a segmented analysis across respondents who use GenAI in research, those who use it only outside research contexts, and those who do not use GenAI.

As most respondents reported using GenAI for research, we present the main figures in the paper for this group to improve readability and focus. Corresponding figures for the other two groups are included in the online replication package.

\subsubsection{Open questions}

\paragraph{Overview}
We applied qualitative analysis to all open-ended questions. For each question, coding was conducted by one author and subsequently reviewed with three other authors, all of whom have extensive qualitative research experience. Disagreements were resolved through negotiated agreement~\cite{campbell2013coding} across two meetings. During this process, researchers discussed the rationale underlying each coding decision until consensus was reached~\cite{garrison2006revisiting}.

\paragraph{Deductive and inductive analysis}
For both RQ2.2 (how is GenAI being used) and RQ3.3 (opportunities), we combined thematic analysis (deductive) \cite{braun2006using} with inductive open coding~\cite{MerriamBook}. 
The deductive phase was grounded in the phase based framework of Andersen et al.~\cite{andersen2025generative}, which categorizes 32 GenAI use cases across five research phases: Idea Generation, Research Design, Data Collection, Data Analysis, and Writing and Reporting. This shared framework served as the analytical scaffold for examining how GenAI use and envisioned opportunities vary across research activities. During analysis, we identified additional use cases and research phases not captured by the original framework. We therefore complemented deductive coding with inductive open coding to incorporate emergent categories.

The question about how is GenAI being used (RQ2.2) was presented only to participants who indicated they use GenAI for SE research and was answered by 151 participants. Among these, 136 described research related use cases and were included in the analysis. 
The remaining 15 responses were excluded because they focused exclusively on teaching related uses (four responses) or described SE practice rather than SE research (for example, \textit{``Help coding organizing documents"}) (11 responses). Of the 136 included responses, 71 mentioned more than one research related use case, resulting in multi label coding and a total of 243 coded use cases.
The questions about using GenAI for SE peer review  (RQ2.3) received 255 responses with 77 explanations about how GenAI is being used for peer review.

The question about envisioned opportunities from using GenAI for SE research (RQ3.3) was answered by 131 participants. One participant referred to a previous response and reported no specific opportunity, and two provided only high-level perspectives, noting that GenAI may make information executable (R15) and enable \textit{``more probabilistic approaches"} (R44). 
We analyzed the remaining 128 responses. Of these, 66 mentioned more than one opportunity, resulting in a total of 266 coded opportunities. We used the same deductive framework, extended with the emergent categories identified during RQ2.2.

\paragraph{Fully inductive analysis}
For RQ4.3 (risk mitigation), RQ5 (need to regulate), RQ2.3 (use in peer review), and the Final Thoughts question, we employed fully inductive open coding~\cite{MerriamBook}, deriving codes directly from participants’ responses.

The question about suggested strategies to mitigate risks from using GenAI for SE research (RQ4.3) was answered by 158 participants. Ten indicated they did not know, eight reported not using GenAI, and two stated that risks could not be mitigated. The remaining 138 responses described at least one mitigation strategy without sacrificing the potential benefits of GenAI.

The question regarding the perceived needs to regulate GenAI in SE research (RQ5) received 119 responses. 

The final open-ended question invited respondents to share any additional thoughts on GenAI in SE research. We performed qualitative coding on the 104 segments from 63 respondents, identifying 36 codes organized into 10 candidate themes.
Since this question was unconstrained in scope, the resulting themes span multiple research questions and cannot be mapped to a single RQ. Therefore, we integrated them as corroborating qualitative evidence within the relevant RQ sections, and synthesized the cross-cutting themes in the Discussion (see Sect.~\ref{sec:discussion}).


\subsection{Replication package}\label{sec:method:replication}
To support transparency and enable replication, we provide a comprehensive replication package publicly available on Figshare.\footnote{\url{https://figshare.com/s/12b873956384863a7c06}} 
The package includes: (i) the complete survey instrument, (ii) the anonymized response dataset with all personally identifiable information removed, and (iii) the codebooks for qualitative analyses.

\section{Results}

In this section, we report our findings structured around the research questions. 

\subsection{Who is using GenAI in SE research, and what is their motivation? (RQ1)}
\label{rq1}

In this research question, we analyze the demographic distribution of SE researchers who reported using GenAI, including gender, career stage, years in current position, organizational affiliation, and geographic distribution. We also analyze the reported motivations to use GenAI.

\subsubsection{Who is using GenAI for their research? (RQ1.1)}
\label{sec:rq1.1}

\begin{figure} [htb]
    \centering
    \includegraphics[width=1\linewidth]{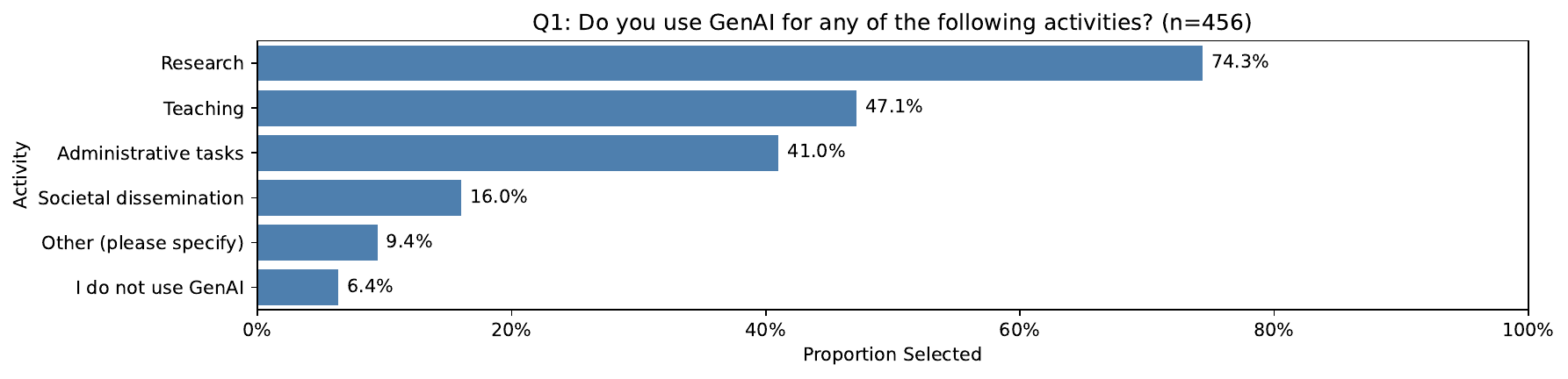}
    \vspace{-6mm}
    \caption{GenAI Activity Distribution}
    \label{fig:Q1_genai_user_groups_distribution}
\end{figure}

Most respondents reported using GenAI for research (74\%), with few reporting using it for non-research purposes (11\%) or not using it at all (7\%) (see Fig.~\ref{fig:Q1_genai_user_groups_distribution}).

\begin{table}[htbp]
\centering
\small
\setlength{\tabcolsep}{6pt}
\renewcommand{\arraystretch}{1.15}
\caption{Demographics of survey respondents. \textit{n} is the number of respondents in each category. \textit{Sample (\%)} is the percentage of respondents in that category relative to all valid responses for that demographic variable (i.e., within each demographic variable, the percentages sum to 100\%). \textit{Use GenAI for Research (\%)} is the percentage of respondents within that category who indicated they used Generative AI for software engineering research (Q1). For Organization, respondents could select multiple options; therefore, the summed counts across its categories can exceed the totals for single-choice demographics.}
\label{tab:demographics}
    \begin{tabular}{llrrr}
        \toprule
        \textbf{Demographics} & \textbf{Category} & \textbf{n} & \textbf{Sample (\%)} & \textbf{Use GenAI for Research (\%)} \\
        \midrule
        \textbf{Continental Origin} & Europe & 111 & 47 & 73 \\
         & North America & 64 & 27 & 84 \\
         & Asia & 41 & 17 & 93 \\
         & South America & 17 & 7 & 65 \\
         & Oceania & 5 & 2 & 60 \\
        \midrule
        \textbf{Gender} & Man & 183 & 72 & 80 \\
         & Woman & 63 & 25 & 73 \\
         & Prefer not to say & 4 & 2 & 50 \\
         & Non-Binary & 3 & 1 & 67 \\
         & Other & 0 & 0 & 0 \\
        \midrule
        \textbf{Career Stage} & Early-career & 105 & 42 & 86 \\
         & Mid-career & 80 & 32 & 79 \\
         & Advanced-career & 62 & 25 & 66 \\
         & Prefer not to say & 4 & 1 & 75 \\
        \midrule
        \textbf{Organization} & University & 221 & 77 & 78 \\
         & Research institute or national lab & 25 & 9 & 88 \\
         & For-profit company & 20 & 7 & 75 \\
         & Government & 7 & 2 & 86 \\
         & Prefer not to say & 5 & 2 & 80 \\
         & Non-profit company & 5 & 2 & 100 \\
         & Other - please specify & 2 & 1 & 50 \\
         & Research funder & 1 & 0 & 0 \\
        \bottomrule
    \end{tabular}
\end{table}


\textbf{Geographic distribution:} From Table \ref{tab:demographics}, we find the geographic composition of respondents across GenAI usage groups. Overall, the sample is largely based in Europe (47\%), followed by North America (27\%) and Asia (17\%), with smaller proportions from South America and Oceania. Looking closer at each continent, we see that the proportion of respondents reporting having used GenAI for research, Asia is the biggest (93\%), followed by North America (84\%) and Europe (73\%), with lower proportions (and sample size) in South America (65\%) and Oceania (60\%).



\textbf{Gender distribution:} The overall sample is predominantly composed of men (72\%), followed by women (25\%), with only a small proportion identifying as non-binary (1\%) or preferring not to disclose the gender (2\%), as seen in Table \ref{tab:demographics}. Within each category, men show the highest inclination towards using GenAI for research (80\%), with women slightly lower (73\%). The remaining categories had lower proportions, but also a very small sample size.


\textbf{Career stage:} Overall, as shown in Table \ref{tab:demographics}, the survey respondents showed a shift toward early-career (42\%), followed by mid- (32\%) and advanced-career (25\%). Looking at the proportion of respondents who used GenAI for their research within each career stage, we find a gradual decline as we go from early- to mid to advanced-career (86\%, 79\%, and 66\% respectively).

\textbf{Organizational affiliation:} As shown in Table \ref{tab:demographics}, the overall sample is mostly composed of respondents affiliated with universities (77\%), while smaller proportions report working in research institutes or national laboratories (9\%) and for-profit companies (7\%), with the rest showing negligible representation. Among these three organizational affiliations, Research institutes or national labs had the highest proportion of respondents who used GenAI for their research (88\%), while universities and for-profit companies responded being somewhat less inclined (78\% and 75\%, respectively).


\subsubsection{What motivates researchers to use GenAI? (RQ1.2)}
\label{sec:rq1.2}

\par \noindent

\begin{figure} [t]
    \centering
    \includegraphics[width=0.7\linewidth]{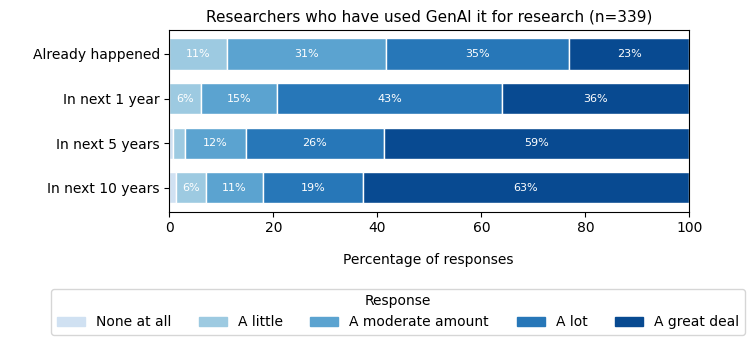}
    \vspace{-3mm}
    \caption{Perceived impact (past and future) of GenAI on SE research}  \label{fig:Q4_perceived_impact_by_genAIusers}
\end{figure}

As shown in Figure~\ref{fig:Q4_perceived_impact_by_genAIusers}, most of the researchers who use GenAI for research perceive its impact as increasingly pronounced in the near to mid-term future. Specifically, 58\% of respondents reported that GenAI has already had a substantial impact (Top-2-Box: \textit{a lot} or \textit{a great deal}), rising to 79\% for the next year and peaking at 85\% for the next five years, before slightly decreasing to 82\% for the next ten years.

\begin{figure} [htb]
     \centering
    \includegraphics[width=1\linewidth]{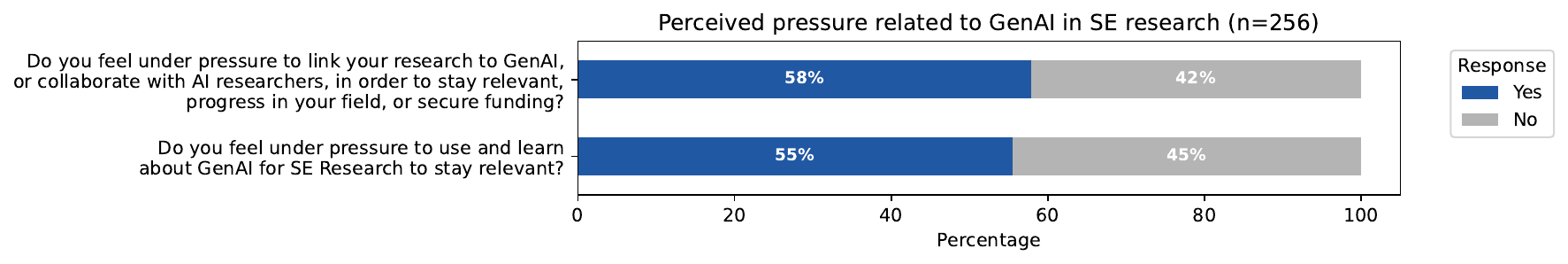}
    \vspace{-6mm}
     \caption{Pressures related to GenAI}
    \vspace{1mm}
    \label{fig:Q12_and_13_GenAI_pressure}
 \end{figure}

\textbf{Pressure related to use and learn about GenAI:} Figure \ref{fig:Q12_and_13_GenAI_pressure} shows how a majority of the researchers feel under pressure to \textsc{link their research to GenAI, or collaborate with AI researchers} (58\%) and also to \textsc{use and learn about GenAI for SE research} (55\%) in order to stay relevant, progress in the field or secure funding.


\textbf{Systemic pressures and persistent skepticism:} The findings above are reinforced by researchers' final thoughts in a last open-text question, which reveal that the pressure to adopt GenAI extends beyond individual motivation to systemic dynamics within academia.
Some respondents described publish-or-perish incentives and funding mandates as key drivers: \textit{``a study without these things is almost automatically considered invalid''} and \textit{``in the project proposal, you MUST mention something about GenAI; otherwise, you're irrelevant''} (R161). 
At the same time, other respondents expressed outright skepticism.
One dismissed GenAI as \textit{``a solution without a problem''} that \textit{``reminds me of Blockchain''} (R306), 
while another expressed concern about the community's direction: \textit{``I feel we have become a second-class AI community''} (R166).


\subsection{Where and how are researchers in SE using GenAI? (RQ2)}
\label{sec:rq2}

This research question provides an overview of how GenAI is being used in SE research, focusing on both \textit{where} it is applied across research strategies, methods, and pipeline stages (Sec.~\ref{sec:rq2.1}), \textit{how} it is used in practice through specific use cases (Sec.~\ref{sec:rq2.2}), and 
\textit{how researchers feel} about the use of GenAI in reviewing research papers (Sec.~\ref{sec:rq2.3}). 


\subsubsection{Where are researchers in SE using GenAI? (research strategy, research method, and research pipeline stage) (RQ2.1)}
\label{sec:rq2.1}

\begin{figure}[tb]
    \centering
    \includegraphics[width=1\linewidth]{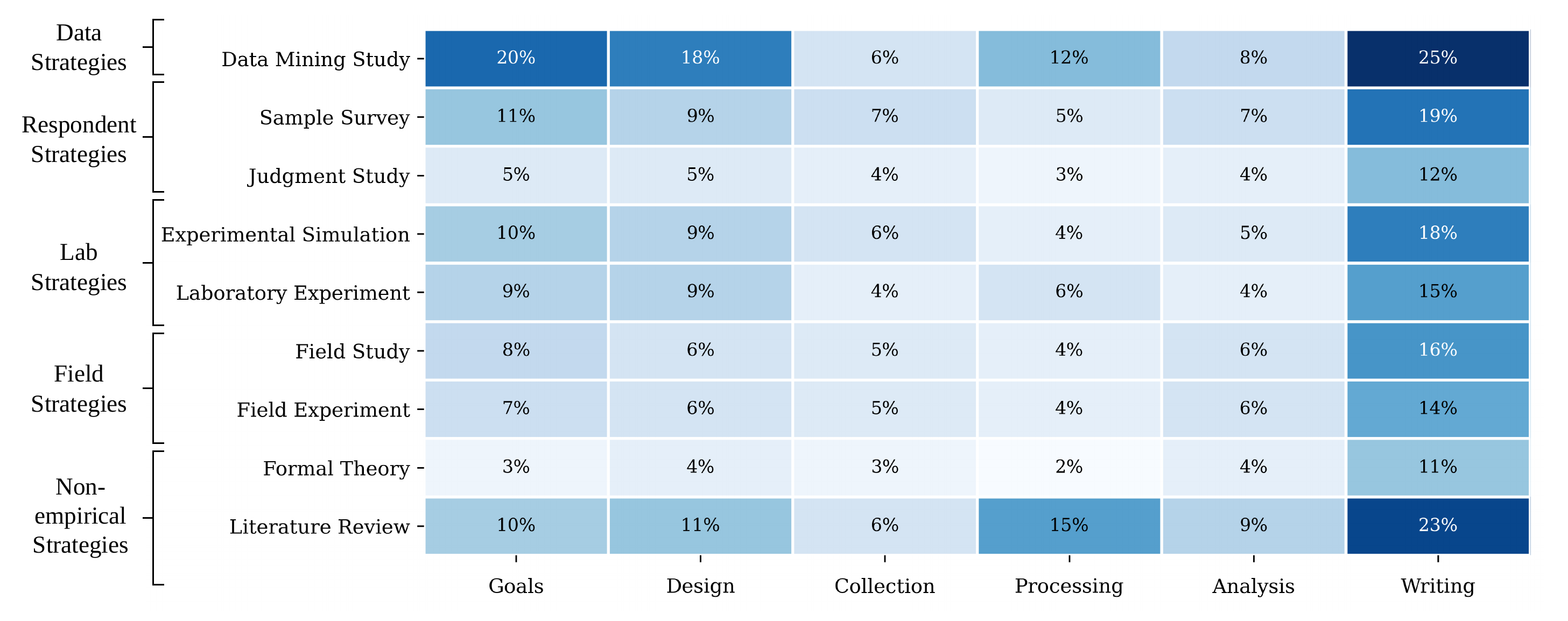}
    \vspace{-6mm}
    \caption{Where GenAI is being used. Research strategies and methods in the Y axis and research pipeline stages in the X axis.}
    \label{fig:Q3_GenAI_for_research_activities}
\end{figure}

We analyze where GenAI is used using the \textit{Who–What–How} framework of Storey et al.~\cite{storey2020software}, focusing on the \textit{How} dimension (research strategies) and the stages of the research pipeline proposed by Trinkenreich et al.~\cite{trinkenreich2025get} (Figure~\ref{fig:Q3_GenAI_for_research_activities}).

GenAI usage is most concentrated in \textbf{data strategies}, with \textit{data mining studies} showing the highest adoption across all pipeline stages. \textit{Respondent strategies} (e.g., surveys, judgment studies) show moderate usage, while \textit{lab} and \textit{field strategies} consistently report lower adoption. Among \textit{non-empirical strategies}, \textit{formal theory} shows the lowest usage, whereas \textit{literature reviews} show relatively high usage, particularly in writing.

Across pipeline stages, GenAI usage is highest in \textbf{writing and dissemination}, followed by early-stage activities such as \textit{goals} and \textit{design}. In contrast, \textit{data collection}, \textit{processing}, and \textit{analysis} show consistently lower usage across strategies.

\subsubsection{How are researchers in SE using GenAI? (RQ2.2)}
\label{sec:rq2.2}

\par\noindent

This research question investigates how SE researchers integrate GenAI into their research practices. Instead of treating GenAI use as a single phenomenon, we analyze how GenAI is used across use cases.

\begin{figure*}[htb]
    \centering
\includegraphics[width=1\textwidth]{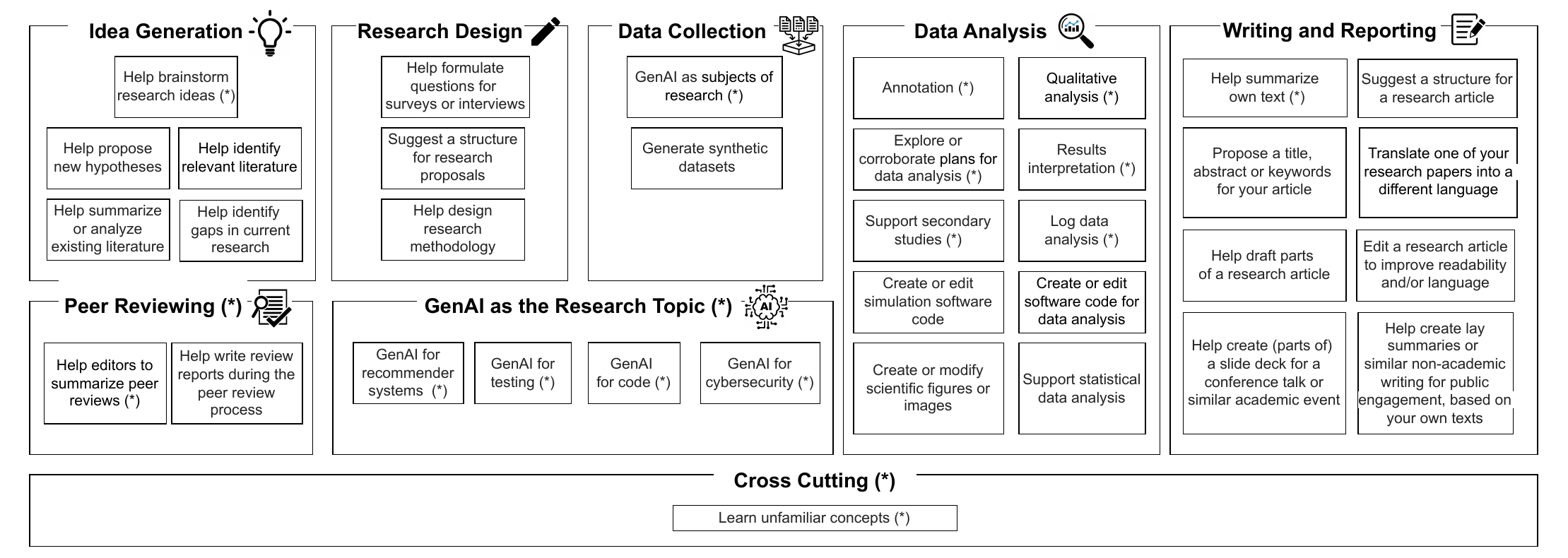}
\vspace{-6mm}
\caption{GenAI use cases categorized using Andersen et al.'s framework~\cite{andersen2025generative}. We mark with (*) the categories and use cases that emerged from our data and are not part of ~\cite{andersen2025generative}.}
\label{fig:usecases}
\end{figure*}

Results are organized using the Andersen et al.~framework~\cite{andersen2025generative}, extended with emergent categories and use cases derived from responses, as shown in Figure~\ref{fig:usecases}.

Table~\ref{tab:GenAIusecases} reports the number of participants whose responses were coded into each category. In the following, we describe these findings in more detail, organized by GenAI use-case category. Because participants’ responses could reflect multiple practices, individual responses were sometimes coded into more than one category.

\begin{table*}[t]
\centering
\small
\caption{Representative examples of answers to how GenAI is being used in SE research, number and percentage of use cases whose answer was coded for each category. We use (**) to represent new categories that are not part of Andersen et al.'s framework  \cite{andersen2025generative}.}
\label{tab:GenAIusecases}
\begin{tabular}{c|p{8cm}|c|r}
\hline
\toprule
 \textbf{Category} & \textbf{Representative examples} & \textbf{\# mentions}  & \textbf{\% (n=243)} \\
 \hline

\toprule
\begin{tabular}[c]{@{}c@{}}Idea \\Generation \end{tabular} & 
 \begin{tabular} [l]{@{}l@{}} \textit{``ask ChatGPT about my research ideas and let it help me} \\ \textit{analyse them"} (R287)\\ \textit{"generate summaries of papers already published"} \\ (R13) \end{tabular} & 52 & 21\% \\ \hline

\begin{tabular}[c]{@{}c@{}}Research \\Design \end{tabular}& \begin{tabular}[l]{@{}l@{}}\textit{``Checking my memory on research methods."} \\(R13)\end{tabular} & 9 & 4\% \\ \hline

\begin{tabular}[c]{@{}c@{}}Data \\Collection \end{tabular}& \begin{tabular}[l]{@{}l@{}}\textit{``as a subject of my research"} (R39) \\\textit{``produce pilot datasets and scenarios to evaluate research} \\\textit{protocols"} (R202)\end{tabular} & 2 & 1\% \\ \hline

\begin{tabular}[c]{@{}c@{}}Data \\Analysis \end{tabular}& \begin{tabular}[l]{@{}l@{}}\textit{``code and analyze interview transcripts and survey responses"} \\(R6), \\ \textit{``making annotations on data [...] and open coding"} \\ (R47)\\ \textit{``help in writing scripts for evaluation purposes"} \\(R75)\end{tabular} & 59 & 24\% \\ \hline

\begin{tabular}[c]{@{}c@{}}Writing \\and Reporting \end{tabular}& \begin{tabular}[l]{@{}l@{}}\textit{``review of research papers that I did; like, possible gaps} \\ \textit{and inconsistencies"} (R71), \\\textit{``when I am writing a report or paper, I seek feedback on the} \\\textit{clarity and accuracy of my language."} (R217) \\\textit{``refine my writing (e.g., grammar checks, clarity of the} \\\textit{sentences)"} (R317)\end{tabular} & 94 & 39\% \\ \hline

\begin{tabular}[c]{@{}c@{}}Peer \\Reviewing (**) \end{tabular}& \begin{tabular}[l]{@{}l@{}}\textit{``revising difficult texts like rejection emails for careful tone"} \\ (R71), \\\textit{``improving text when writing [..] reviews"} \\ (R167) \end{tabular} & 4 & 2\% \\ \hline

\begin{tabular}[c]{@{}c@{}}GenAI as \\Research Topic (**) \end{tabular}& \begin{tabular}[l]{@{}l@{}}\textit{``experimenting with the usage of GenAI in test case and UML} \\\textit{models generation"} (R46), \\\textit{``examining use of LLMs for generating code"} \\(R82) \\\textit{``1) Secure coding; 2) explanation of ransomware attack strategy;} \\\textit{3) identifying the pattern of use of GenAi for secure coding} \\\textit{tasks"} (R278)\end{tabular} & 12 & 5\% \\ \hline

\begin{tabular}[c]{@{}c@{}}Cross \\Cutting (**) \end{tabular}& \begin{tabular}[l]{@{}l@{}}\textit{``checking concepts that I am unfamiliar with"} \\(R186), \\\textit{``understand new technologies, new concepts"} \\(R152) \\\textit{``Help with learning/using APIs and other programming} \\\textit{features"} (R28)\end{tabular} & 11 & 4\% \\ \hline

\bottomrule

\end{tabular}

\begin{center}
\parbox{0.9\textwidth}{%
\footnotesize
The total per category is not the sum of the respondents since participants often provided an answer that was categorized into \textbf{more than one use case} (e.g., Idea Generation and Data Analysis).
}
\end{center}

\end{table*}


In the \textbf{idea generation} phase, GenAI is predominantly used to support early-stage sensemaking and orientation activities. Researchers mentioned using GenAI to \textsc{help brainstorm research ideas}, either in the \textit{``very early stages [..] for a quick overview of the state of the practice"} (R314) or \textit{``not for initial brainstorming, but to refine ideas"} (R18). In this role, GenAI was described as \textit{``a phenomenal tool for reflecting on new ideas, as often and as deeply as [they] need"} (R2). 
The brainstorm sometimes leads to \textsc{help propose new hypotheses,} and respondents mentioned to use GenAI for \textit{``tun[ing] the questions and hypotesis"} (R336) to \textit{``find a better formulation or statements"} (R379).

Beyond brainstorming and support on hypotheses, respondents reported using GenAI to \textsc{\textit{``help identify relevant literature"}} (R75) as a way to \textit{“quickly learn or obtain summaries about specific topics”} (R85). This included support for finding references (R69, R199) and related work (R66, R70, R206, R292), for example, by \textit{``using deep research"} (R86, R426), or more generally, obtaining \textit{``basic overviews of topic areas"} (R265). Still focusing on literature, respondents reported using GenAI to \textsc{help summarize or analyze existing literature}, including articles (R335) as well as different types of \textit{``texts, videos and audios"} (R52). Beyond summarization, GenAI was also used to \textit{``explain [others] papers"} (R426) and to \textsc{help identify gaps in current research} by \textit{``anticipating potential reviewer interpretations of specific sentences or results."} (R84).


As researchers transition into the \textbf{research design} phase, the use of GenAI was mentioned with more restraint. Here GenAI is primarily employed to \textsc{help design research methodology}, particularly by assisting with the clarification and examination of methodological steps. This is evidenced by reports of using GenAI to \textit{"discuss methodology steps and possible associated threats"} (R303) and for  \textit{"clarifying some steps of research methods"} (R264). Another related use case involves methodological exploration, where GenAI is used to investigate \textit{“whether a more ‘natural’ or intuitive approach exists for a given problem” }(R84). In addition to these methodological supports, some respondents report using GenAI to assist with the \textsc{suggest a structure for research proposals}, such as generating \textit{“skeletons for documents and grant applications” }(R80).


In the \textbf{data collection} phase, references to GenAI use appear less frequently and are mainly associated with supporting and exploratory activities. When GenAI is mentioned in this phase, it is not described as a mechanism for directly collecting empirical data, but rather as a complementary resource within the data collection process. In this context, some respondents refer to \textsc{GenAI as subjects of research}, describing its use \textit{“as a subject of [their] research”} (R39). Others report employing GenAI to \textsc{Generate synthetic datasets}, using it to \textit{“produce pilot datasets and scenarios to evaluate research protocols”} (R202).


The \textbf{data analysis} phase represents one of the most technically grounded areas of GenAI use. Researchers frequently mentioned using GenAI to augment analytical labor, particularly for creating or editing analysis code, supporting exploratory data analysis, and assisting qualitative analysis. A recurring pattern involves delegating mechanical or repetitive tasks to GenAI while retaining human oversight over interpretation and validation.

 Respondents describe using GenAI to support \textsc{annotation} activities as part of the analytical process. One participant notes that \textit{"have also experimented with its use for annotations. For instance, [he] asked a student to annotate a dataset and then prompted ChatGPT to do the same. Thereby, [they] spotted overlooked cases and computed the corresponding inter-rater reliability score"} (R377). In \textsc{qualitative analysis} contexts, GenAI is similarly employed to assist with early analytical steps, including \textit{"support the analysis of qualitative data"}(R157) and \textit{"making annotations on data [..] and open coding"} (R47).

 GenAI is used to \textsc{explore or corroborate plans for data analysis}, including to \textit{"verify whether some abductive lines of thought can be grounded in some anecdotal evidence"} (R85). It is also employed during \textsc{results interpretation}, where respondents describe using it to \textit{"analyze, interpret and make sense of my data"} (R317) and to \textit{provide possible interpretations of my results} (R44).

Respondents also report using GenAI to \textsc{support secondary studies}, such as \textit{support clerical activities in secondary studies} (R202). \textsc{Log data analysis} is another reported use, including \textit{"analyzing log data, detecting anomalies"} (R141), as well as the \textsc{create or modify scientific figures or images}, such as \textit{"graph generation"} (R60).

GenAI is frequently used to \textsc{create or edit software code for data analysis}. Participants report using it to \textit{"generate the code for my research experiments"} (R317), assist with \textit{"training parameters, polishing the prompts"} (R160), and integrate it \textit{"as part of my toolbox for analyzing data"} (R73). It is also used to \textsc{support statistical data analysis}, such as \textit{"selecting the right statistical tests for a certain purpose"} (R160).


The \textbf{writing and reporting} phase was the most mentioned one (see Table \ref{tab:GenAIusecases}). Core use cases focus on linguistic and communicative support, including improving readability, rephrasing text, and summarizing one’s own writing. \textsc{Help summarize own text} appears in practices such as \textit{"summarizing notes"} (R206) and \textit{"getting synopsis of (old) lines of work"} (R69).  Relatedly, GenAI is used to \textsc{suggest a structure for a research article}, for example, through \textit{"structuring text from bullets"} (R43). It is also employed to \textsc{propose a title, abstract or keywords for [the] article}, including \textit{"brainstorming paper titles"} (R47) and \textit{"give ideas (e.g., of paper titles)"} (R134), as well as to \textsc{translate one of your research papers into a different language}, such as \textit{"language translation"}(R2).

In addition, respondents report using GenAI to \textsc{help draft parts of a research article}, including \textit{"refine paper"} (R19) and \textit{"review of research papers that I did; like, possible gaps and inconsistencies"} (R303). GenAI is also used to \textsc{edit a research article to improve readability and/or language}, with examples such as \textit{"when I am writing a report or paper, I seek feedback on the clarity and accuracy of my language"} (R217), \textit{"polishing my writing"} (R186), \textit{"refine my writing (e.g., grammar checks, clarity of the sentences)"} (R317) and \textit{"text reviewing for grammar and fluency"} (R207). Further uses include \textsc{help create (parts of) a slide deck for a conference talk or similar academic event}, such as \textit{"urning lecture transcripts into prose texts"} (R71), and \textsc{help create lay summaries or similar non-academic writing for public engagement, based on your own texts}, for example \textit{"creating a summarized version for social media"} (R314).

In the \textbf{peer review} phase, GenAI use is framed as supportive of comprehension and communication rather than evaluation. Respondents note that GenAI can \textsc{help write review reports during the peer-review process}, particularly for tasks that require careful wording and tone. This includes uses such as \textit{"revising difficult texts like rejection emails for careful tone"} (R71), as well as \textit{"checking communications and paper reviews"} (R215).


In the \textbf{cross-cutting} phase, the use of generative AI is not confined to a single stage of the research lifecycle, but instead spans multiple activities across different phases. From this perspective, participants describe using GenAI to support ongoing learning and orientation, particularly to \textsc{learn unfamiliar concepts}, such as \textit{“checking concepts that I am unfamiliar with”} (R186) and \textit{“learn about some specific techniques”} (R16).

Finally, when \textbf{GenAI itself is the research topic}, responses describe cases in which researchers explicitly study GenAI-enabled techniques as the primary object of investigation. These studies span different use cases of Software Engineering, including security, recommender systems, and testing. 


\subsubsection{How do researchers feel about the use of GenAI in reviewing research papers? (RQ2.3)}
\label{sec:rq2.3}

\par\noindent

In the survey we asked an open-ended question for those that answered yes to: ``As a reviewer, have you used GenAI to assist in reviewing a SE research paper?''.

Eighty-one respondents shared that they used GenAI to support reviewing activities with most explaining how they use it. Most explained they found GenAI support for reviewing useful, but five participants mentioned they didn't find it helpful, and three had mixed experiences using GenAI during reviewing. Notably one participant mentioned using it to review their own paper.

Below, we present how respondents use GenAI to formulate their reviews, to check the paper against paper acceptance criteria, to provide cognitive support as they conduct their reviews, and the experiences the respondents reported while reviewing using GenAI as a support. Throughout, we discuss the implications of these findings and refer to additional comments the respondents provided as part of the last question in the survey for final thoughts.  Finally, we share some additional insights respondents shared about their view of how GenAI is used in reviewing in our community.

\begin{figure}[t]
\centering
\includegraphics[width=1\textwidth]{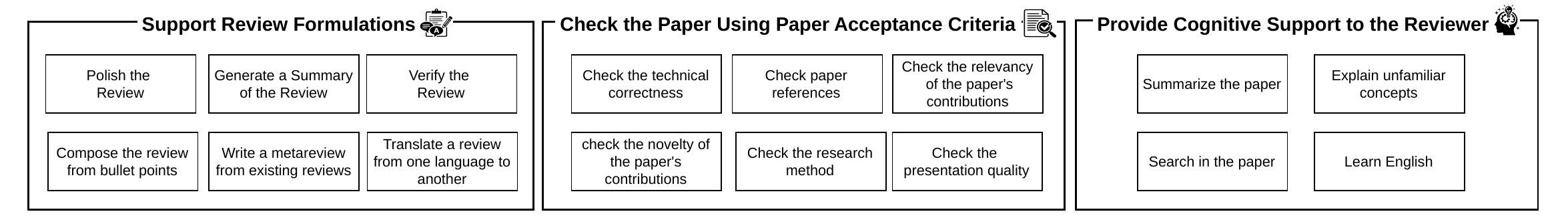}
\caption{The Use Cases of GenAI for Peer Review.}
\label{fig:peerreview}
\end{figure}

\paragraph{GenAI is Used to Support Review Formulations}

Many of the participants explained how they use GenAI to support them in formulating their reviews. Figure \ref{fig:peerreview} summarizes the themes that emerged for how GenAI supports review formulation. 

Forty-one participants mentioned they use it to \textbf{polish their review} only: \textit{``I use GenAI to rephrase and refine my review comments. This helps improve clarity and tone, making feedback more understandable and constructive for authors and fellow reviewers''}(R23). While another described how they use GenAI to \textbf{translate their review} from one language to another: \textit{``to write my feedback in English as a translator"}(R411). Seven respondents described how they use GenAI to \textbf{verify their review}: \textit{``I submit both the paper an my review and ask the model to challenge the review. Then I *critically* analyze the response.''} (R219) Three mentioned that they use GenAI to \textbf{summarize the main points in their review}: \textit{``Only to summarize strengths and weaknesses of a paper based on my detailed comments''} (R65). One participant mentioned how they use GenAI to \textbf{write the review}  from their bullet points: \textit{``Write reviews based on my bullet points from reading the paper; check for additional strengths and weaknesses that I overlooked; improve grammar and writing,''} (R310) while another mentioned using GenAI to help them \textbf{write a metareview} 
from existing reviews: \textit{``Once I used it to write meta-reviews summarizing the existing reviews''} (R43).



In the final question of the survey that asked respondents about their final thoughts, several respondents corroborated the view that GenAI use in reviews should be limited to rephrasing. 
One respondent argued that this type of \textit{``GenAI-Rewriting-Confirmation should be allowed in any written communications, including paper and peer review writing''} (R34). R332 drew the line explicitly, stating that reviewers should \textit{``only be allowed to use it to rewrite what they have already written,''} consistently with the existing ACM policy on authorship.~\footnote{\url{https://www.acm.org/publications/policies/new-acm-policy-on-authorship}}

\paragraph{GenAI Used to Check the Paper Using Paper Acceptance Criteria}

Several respondents described how they use GenAI to help them review papers using acceptance criteria that are often used for paper reviews in software engineering. Four respondents mentioned using GenAI to \textbf{check the technical correctness} of the paper. For example, R343 said: \textit{``By selecting and pointing out all the logical inconsistencies and shortcomings in the publication.''} Two mentioned using GenAI to \textbf{check paper references}:  \textit{``assisting with reference and data checks''}
(R327).

Two respondents mentioned using GenAI to help them \textbf{check the relevancy of the paper's contributions}: For example, \textit{``I asked GenAI to estimate the top X factors and challenges of an SE phenomenon because I thought the paper's results were obvious. GenAI produced the same results as the paper but with better insight. I agreed and recommended rejection of the study as `obvious''}(R70). One participant also used GenAI to \textbf{check the novelty of the paper's contributions}: \textit{``To help search for related literature and double check the novelty of contribution.''}(R257)

Two respondents mentioned they used GenAI to \textbf{check the research method} used for the research in the paper under review. For example, \textit{``I used it only to verify whether a design method is suitable for the research goal at hand when I am not much experienced with it.''} (R76) Two respondents mentioned using GenAI to \textbf{check the  presentation quality} of the paper. For example, \textit{``checking writing of a part of the article where I think is a problem with English. Inspecting availability of tools and references''}    (R247).

The final thoughts open-text question revealed strong views on how far such analytical use should extend. Several respondents stated that GenAI may be acceptable for reading the submission, but is \textit{``questionable for detecting pros and cons, and is inacceptable for making a decision''} (R41). Similarly, one respondent argued that GenAI \textit{``should NOT be given the paper itself and asked what it thinks''} and should instead be limited to Grammarly-like functions (R48). Another drew the line at content analysis, stating that reviewers should be \textit{``allowed to use GenAI only to rephrase the text, and it should be forbidden [to] use [it] to analyze a paper, review the literature or check the results/code''} (R184).

\paragraph{GenAI Provides Cognitive Support to the Reviewer}

In addition to using GenAI to support review writing and review formulation, several respondents described how they used GenAI to provide other types of cognitive support as they were reviewing a paper.  These additional types of cognitive support that emerged from the open-ended responses are summarized next. Six researchers described how they use GenAI to \textbf{summarize the paper} to aid their understanding of the paper:  \textit{``At beginning to get quick summary of the paper''} (R193). An additional two respondents describes how they use GenAI to \textbf{explain unfamiliar concepts} or to provide a summary of an unfamiliar topic: \textit{``...and also to clarify concepts I am not familiar with''} (R214). One respondent described how they used GenAI to \textbf{search in the paper} to \textit{``...find in some case the link of the online appendix''} (R45). And one participant mentioned how they use GenAI to help them \textbf{learn English} as they are reviewing, not just to improve the review: \textit{``To improve my English and writing skills for my final revision. Again, I use it for more grammatical aspects and to see if what I want to say is being conveyed correctly''} (R435).

\paragraph{Poor Experiences Using GenAI for Reviewing}

Although, respondents described the many ways GenAI supported them while reviewing, not all had positive experiences.  For example, \textit{``I've tried it to see if I missed anything I should.  I've found that it's *really* bad at reviewing papers.  It found critiques of things that were trivial (formatting of the bibliography) while missing fundamental problems in a paper. Reviewing takes higher-level thinking, and my experience (given, this is like an n of 3) is that it lacks that capability.  Maybe the new "thinking" models like O3 would do better.  I'd like it to helpfully review my own papers before submission but I have yet to get a whole lot of utility out of it''} (R7). 


\paragraph{Stances on GenAI Use in Peer Review}

Beyond specific use cases, in the final thoughts question of the survey, 15 respondents spontaneously suggested normative positions on whether GenAI should be permitted in peer review at all, ranging from outright prohibition to conditional usage with safeguards. The most restrictive stance called for a complete ban, even on text polishing. One respondent argued that \textit{``arguments for or against a paper should be of one['s] own''} and that \textit{``not having very-well written reviews do not affect the final quality of the paper,''} concluding that GenAI should \textit{``be forbidden at all (even for polishing reviews)''} (R26). Respondent R315 was equally firm: \textit{``It should not be allowed, even with guidelines. Peer review is a critical quality gate of scientific publications''} and should rely on the ability of peers to \textit{``understand the content of the paper and evaluate its fitness for the target venue.''}

Some respondents framed the issue in terms of professional trust. One stated that reviewers caught using GenAI to write reviews \textit{``shouldn't be welcome in the community going forward, full stop,''} comparing the practice to \textit{``letting a random friend at a bar write your reviews for you''} (R42). Another questioned the very purpose of peer review under GenAI adoption: \textit{``Why have reviewers if they are going to [use] AI?''} (R59).

An intermediate position acknowledged the practical pressures on reviewers while expressing doubt about responsible use in practice. One respondent suggested that \textit{``reviewers should not be encouraged to use it, but allowed to use it with guidelines,''} noting that \textit{``the reviewing load is high and the research community needs many, high-quality reviews,''} but also adding: \textit{``I am very doubtful that people can use it correctly''} (R107). Another respondent called for allowing GenAI \textit{``given how much time it can save''} while stressing that \textit{``this heightens the responsibility of reviewers''} (R82). 

A constructive alternative was proposed by respondent R37 who stated they \textit{``out of principle never use AI for reviewing''} and consider it a duty to report reviewers who misuse GenAI when evaluating assignments, but could also envision \textit{``mixed formats, where a paper is reviewed by humans AND AI in parallel, as a `4th reviewer' so to speak.''} Another respondent envisioned using GenAI to \textit{``generate template-based paper summaries that could be used for a `first round' review of papers,''} while also cautioning that \textit{``there is a risk that the reviewers will also use GenAI in the second round''} (R181).

Finally, respondents raised specific risks tied to the reviewing context. 
One noted that \textit{``the very second I upload a document to GenAI, I potentially breach non-disclosure agreements (also in context of peer review)''} (R158),
highlighting the confidentiality implications in reviewing activities.

\subsection{What benefits, challenges, and opportunities do SE researchers perceive when using GenAI? (RQ3)}
\label{sec:rq3}

In this research question, we analyze the reported benefits, challenges, and opportunities of using GenAI for SE research.

\subsubsection{Which benefits do SE researchers perceive GenAI offers to SE Research? (RQ3.1)}
\par
\begin{figure}[t]
    \centering
    \includegraphics[width=1\linewidth]{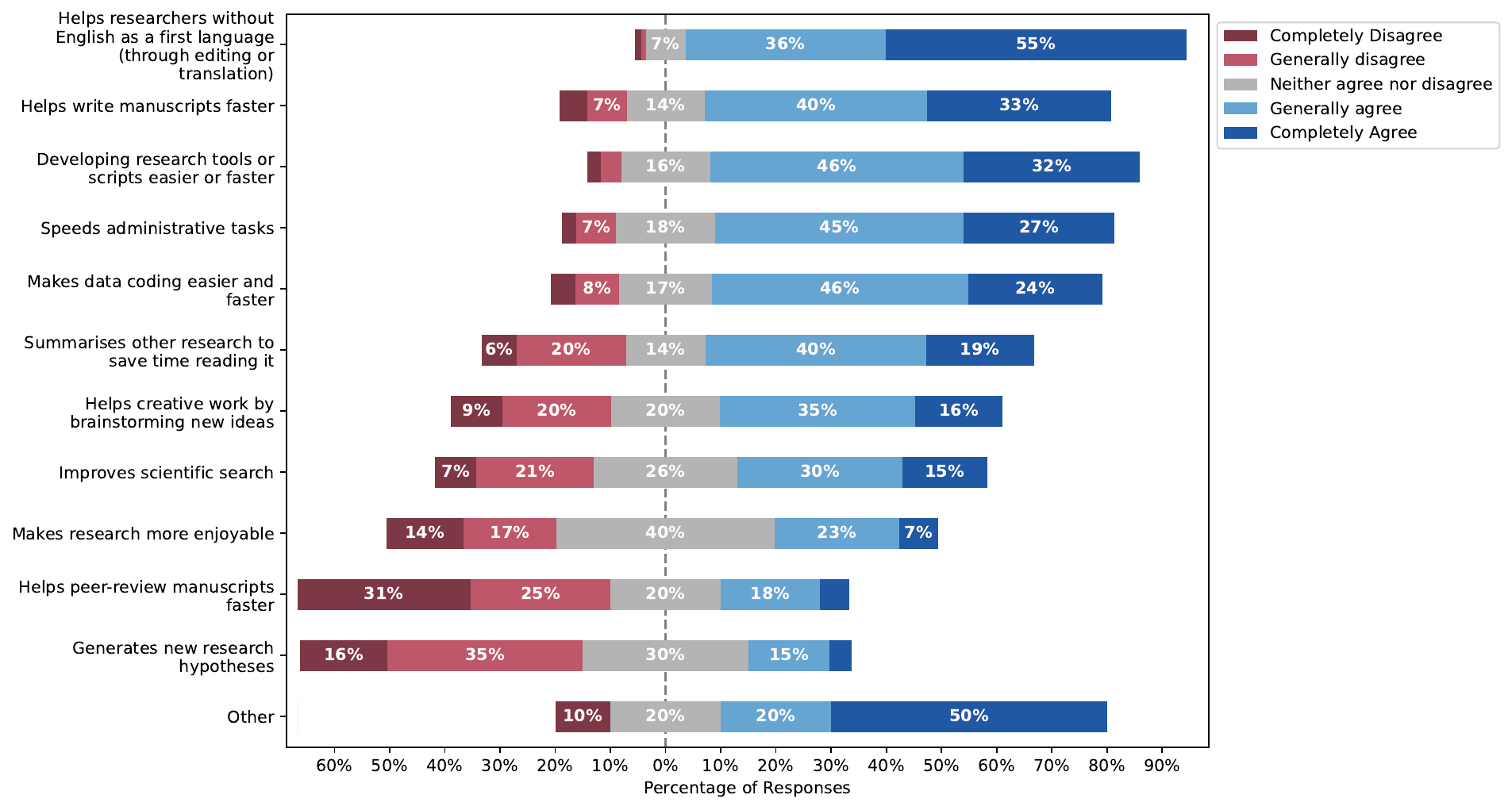}
    \caption{Perceived benefits of using SE research}
    \label{fig:Q5_GenAI_benefits}
\end{figure}

 \noindent
Figure~\ref{fig:Q5_GenAI_benefits} presents the perceived benefits of using GenAI in SE research. To facilitate interpretation, we aggregate \textit{completely disagree} with \textit{generally disagree}, and \textit{generally agree} with \textit{completely agree}. 

In general, the responses indicate a predominantly positive perception of most listed use cases, particularly those related to writing and text processing. The strongest agreement is observed for \textsc{helping non-native English speakers} (91\% agreement), followed by \textsc{easier or faster developing research tools and scripts} (78\%), also \textsc{writing manuscripts faster} (73\%), \textsc{speeding up administrative tasks} (72\%), \textsc{making data coding easier or faster} (70\%), \textsc{summarizing text} (59\%), and \textsc{improving scientific search} (45\%). 

More mixed perceptions emerge for creativity-related activities. While a majority still agrees that GenAI helps \textsc{brainstorming new ideas} (51\%), responses are more distributed, and perceptions are even more neutral for \textsc{making research more enjoyable} (30\% agreement, 40\% neutral). We expand on the opportunities to improve researcher experience in Section \ref{sec:rq3.3}.

In contrast, perceptions are predominantly negative for \textsc{speeding up peer review} (56\% disagreement) and \textsc{generating new research hypotheses} (51\% disagreement).

Researchers' final thoughts reinforced these findings, revealing nuances in how benefits are perceived in practice. 
Regarding text processing (the category with the strongest agreement), one respondent described GenAI as \textit{``excellent in formulating perfect texts [\ldots] great for dissemination results and paraphrasing own's texts''} (R343). 
Beyond text processing, productivity benefits were broadly recognized, though respondents typically coupled them with calls for caution. Respondents noted that GenAI \textit{``definitely helps SE researchers a lot in our daily tasks''} but stressed the need to \textit{``be very careful and establish guidelines to use it''} (R139).
Similarly, while acknowledging that GenAI is \textit{``definitely a tool to boost productivity,''} one respondent underlined that \textit{``we must use it wisely,''} raising concerns that uncritical adoption \textit{``may lead to an explosion of papers, not necessarily of high quality''} (R135).
Others framed GenAI as \textit{``both a transformative opportunity and a serious responsibility,''} emphasizing that \textit{``its use must be grounded in transparency, reproducibility, and ethical standards''} (R23). 
This recurring ``\textit{yes, but}'' pattern where benefits are acknowledged alongside calls for caution, regulation, or verification suggests that even among researchers who perceive clear advantages, uncritical adoption is not endorsed.
We further analyze these conditions and perceived risks in Section~\ref{sec:rq4}.


\subsubsection{What challenges do SE researchers experience using GenAI in SE research? (RQ3.2)}
\par\noindent

\begin{figure}[!t]
         \centering
         \includegraphics[width=1\linewidth]{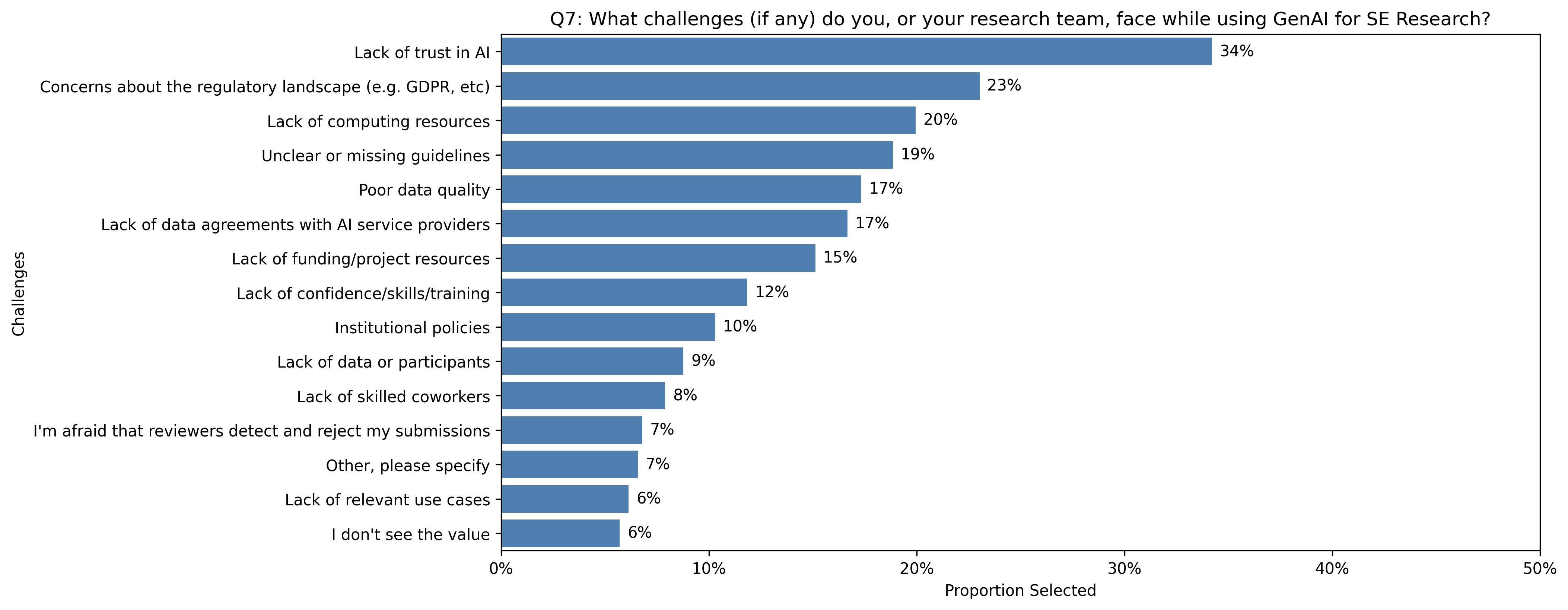}
         \vspace{-6mm}
         \caption{Challenges faced when using GenAI}
         \label{fig:Q7_challenges}
     \end{figure}

Figure \ref{fig:Q7_challenges} shows that overall \textsc{lack of trust} in AI was the most frequently cited challenge, reported by 34\% of respondents. We provide a more detailed analysis of trust-related responses in Section \ref{sec:rq4.3}. The second most commonly reported challenge, mentioned by 23\% of respondents, concerns the \textsc{regulatory landscape} (e.g., GDPR). We examine these regulatory concerns in greater depth in Section \ref{sec:rq5}.
    
\subsubsection{What opportunities do SE researchers envision GenAI offers to SE Research? (RQ3.3)}
\label{sec:rq3.3}
\par\noindent

\begin{figure*}[!bth]
\centering
\includegraphics[width=1\textwidth]{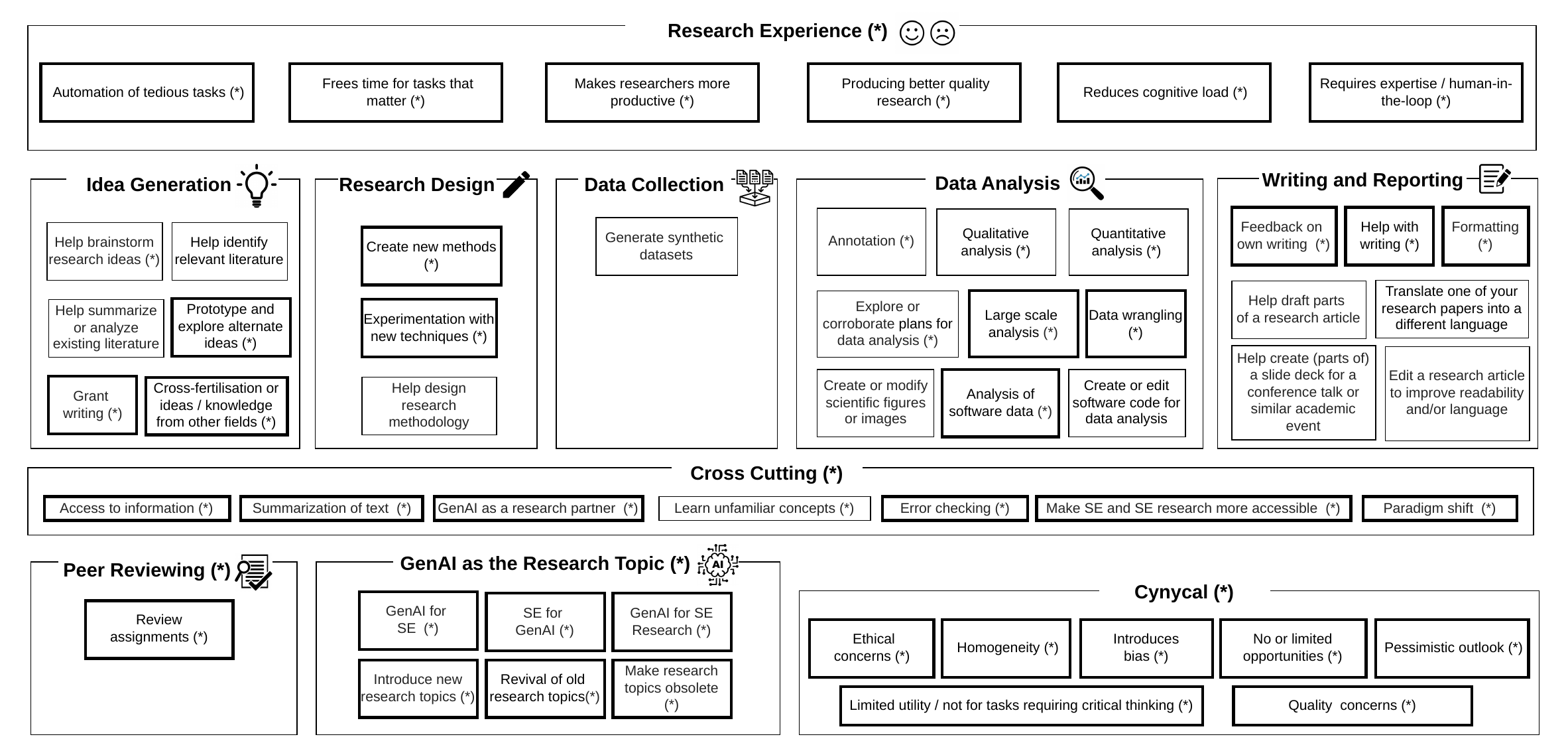}
\vspace{-6mm}
\caption{GenAI opportunities categorized using Andersen et al.'s framework~\cite{andersen2025generative}. We marked as (*) the categories and opportunities that emerged from our data and were not part of ~\cite{andersen2025generative} and bold boxes as the opportunities that were not mentioned as current use cases in Section \ref{sec:rq2.2}.}
\label{fig:opportunities}
\end{figure*}

This research question investigates which GenAI use cases are envisioned as \textsc{opportunities} for SE research. Results are organized using the Andersen et al.~framework~\cite{andersen2025generative}, extended with emergent categories derived from participants’ responses, as illustrated in Figure~\ref{fig:opportunities}. Our qualitative analysis identified 15 opportunity categories that overlap with current GenAI use cases reported in Section~\ref{sec:rq2.2}. In Figure~\ref{fig:opportunities}, opportunity categories \textbf{not} previously observed as current uses are highlighted in bold and described below. 

Table~\ref{tab:GenAIopportunities} summarizes the number of participants whose responses map to each category. We next present detailed findings organized according to these categories of GenAI opportunities for SE research. Because participants’ responses could reflect multiple opportunities, individual responses were sometimes coded into more than one category.

\begin{table*}[!tbh]
\centering
\small
\caption{Representative examples of opportunities GenAI bring to SE research, number and percentage of use cases whose answer was coded for each category. We use (**) to represent new categories that were not part of Andersen et al.'s framework  \cite{andersen2025generative}}
\label{tab:GenAIopportunities}
\begin{tabular}{c|p{8cm}|c|r}
\hline
\toprule
 \textbf{Category} & \textbf{Representative examples} & \textbf{\# mentions}  & \textbf{\% (n=266)} \\
 \hline

\toprule
\begin{tabular}[c]{@{}c@{}}Idea \\Generation \end{tabular} & 
 \begin{tabular} [l]{@{}l@{}} \textit{``can be great to kickstart a literature search, summarize papers}\\\textit{to see if they are relevant and warrant a deeper look, and similar}\\ \textit{tasks"} (R63) \end{tabular} & 37 & 14\% \\ \hline

\begin{tabular}[c]{@{}c@{}}Research \\Design \end{tabular}& 
\begin{tabular} [l]{@{}l@{}} \textit{``Faster experimentation with techniques that a researcher is not}\\\textit{familiar with by specifying the intended research goal as a}\\ \textit{prompt"} (R201) \end{tabular} & 8 & 3\% \\ \hline

\begin{tabular}[c]{@{}c@{}}Data \\Collection \end{tabular}&
\begin{tabular} [l]{@{}l@{}} \textit{``generating synthetic data"} (R187) \end{tabular} & 7 & 3\% \\ \hline

\begin{tabular}[c]{@{}c@{}}Data \\Analysis \end{tabular}&
\begin{tabular} [l]{@{}l@{}} \textit{``generate quick tools and scripts for analyzing data"}\\ (R82) \end{tabular} & 22 & 8\% \\ \hline

\begin{tabular}[c]{@{}c@{}}Writing \\and Reporting \end{tabular}&
\begin{tabular} [l]{@{}l@{}} \textit{``Help the researcher to go through multiple documents in less}\\ \textit{time and express the ideas in better English"} \\(R340) \end{tabular} & 19 & 7\% \\ \hline

\begin{tabular}[c]{@{}c@{}}Peer \\Reviewing (**) \end{tabular}& 
\begin{tabular} [l]{@{}l@{}} \textit{``GenAI can help with a fairer allocation of paper review tasks}\\ \textit{for SE conferences based on reviewers' prior publications"} \\(R17) \end{tabular} & 1 & 0\% \\ \hline

\begin{tabular}[c]{@{}c@{}}GenAI as \\Research Topic (**) \end{tabular}&
\begin{tabular} [l]{@{}l@{}} \textit{`explor[ing] how humans and AI systems co-develop software,}\\ \textit{raising new questions in usability, trust, and explainability of} \\ \textit{AI-generated artifacts"} (R23) \end{tabular} & 15 & 6\% \\ \hline

\begin{tabular}[c]{@{}c@{}}Cross \\Cutting (**) \end{tabular}& 
\begin{tabular} [l]{@{}l@{}} \textit{``GenAI can serve as a consultant throughout a research project"} \\(R174) \\ \textit{``as an additional author [..] to reduce researcher bias and} \\ \textit{highlight areas that are overseen by the authors"} \\(R215) \end{tabular} & 23 & 9\% \\ \hline

\begin{tabular}[c]{@{}c@{}}Cynical (**) \end{tabular}& 
\begin{tabular} [l]{@{}l@{}} \textit{``Ethical concerns arise [..] and some misconduct in paper writing}\\ \textit{and reviews have already been seen [..] [raising] a blurry line [..]}\\\textit{on the edge of ethical principles"} (R165) \end{tabular} & 29 & 11\% \\ \hline


\begin{tabular}[c]{@{}c@{}}Researcher \\Experience (**) \end{tabular}& 
\begin{tabular} [l]{@{}l@{}} \textit{``Similar to when Google freed me from remembering all facts} \\ \textit{about the world, I no longer have to be an expert in a language} \\\textit{or method to be able to use it."} (R156)\\ \textit{``smaller but time-consuming tasks can be sped up"} \\(R63) \end{tabular} & 68 & 25\% \\ \hline

\bottomrule

\end{tabular}

\par\vspace{0.3em}
\noindent
\parbox{0.9\textwidth}{%
\footnotesize
The total per category is not the sum of the respondents since participants often provided an answer that was categorized into \textbf{more than one use case} (e.g., Idea Generation and Data Analysis). 
}

\end{table*}


The most frequently cited opportunities concerned \textbf{researcher experience}, highlighting how GenAI may reshape the day-to-day practice of SE research. A dominant theme was the \textsc{automation of tedious tasks}, including routine coding, labeling, scripting, and other repetitive activities that are \textit{``rather routine human-labor intensive but not ‘intellectually exciting’ aspects of research"} (R56). Such automation was widely associated with increased efficiency and the ability to shift effort toward higher-value activities. Closely related, respondents emphasized that GenAI \textsc{frees time for tasks that matter}. By removing implementation and formatting burdens, researchers can \textit{``focus on the problem and solution rather than on the implementation and dissemination"} (R62) and devote more attention to \textit{innovation and deep thinking for better science} (R164).

Many responses framed these changes as \textsc{making researchers more productive}. Reported benefits included faster prototyping, quicker data analysis, accelerated writing, and the ability to \textit{``do more research within the same limited time"} (R34). Others described broader efficiency gains across the research lifecycle, including speedups in nearly all research steps and fewer mistakes through automated checking (R308).

Beyond productivity, respondents also associated GenAI with \textsc{producing better quality research}. Some highlighted increased output quality and quantity (R138), while others connected automation and intelligence in SE workflows to improved software quality and modeling capabilities (R142). Several responses further indicated a \textsc{reduction in cognitive load}. Examples included no longer needing deep expertise in specific languages or tools to apply them effectively (R156) and mitigating fatigue-related limitations in SE work (R80).

Finally, respondents repeatedly stressed that meaningful benefits still \textsc{require expertise and human-in-the-loop oversight}. GenAI was described as \textit{``a great helper if and only if you already have a fundamental understanding"} (R63), with methodological decisions and interpretation remaining inherently human responsibilities (R336; R380).





In the \textbf{idea generation} phase, respondents envisioned opportunities that were not reported as part of current use (marked with a star in Fig.~\ref{fig:opportunities}). Among these, \textsc{\textit{Grant writing}} was briefly mentioned (R52). Another emerging opportunity concerned the ability to \textsc{prototype and explore alternate ideas}, primarily enabled by faster prototype development (R6; R83; R130). This increased speed was described as allowing researchers to \textit{``explore multiple alternatives and pick the best one"} (R7).

Respondents also envisioned opportunities to \textit{\textsc{cross-fertilize between different research fields}, such as better use of statistical methods and tools"} (R65). In this context, GenAI was described as enabling SE researchers to \textit{broaden research scope by quickly grasp[ing] a new area, technology or concept"} (R153), at least capturing \textit{``the essence"} of unfamiliar domains (R288).


Within the \textbf{research design} phase, respondents envisioned opportunities 
to \textsc{create new methods}, noting that \textit{``the nature of user studies might change in some cases, with the right protocol, it might be easier to automate"} (R85). In addition, GenAI was seen as enabling \textsc{experimentation with new techniques}(R201).





In the \textbf{data analysis} phase, respondents envisioned advances in \textsc{quantitative analysis} (R7, R187) and, relatedly, \textsc{large-scale analysis}, emphasizing support for \textit{big data analysis"} (R326). Respondents further highlighted opportunities for \textsc{data wrangling}, including \textit{optimization of data"} (R152) to \textit{make data preprocessing much easier"} (R333). Finally, opportunities for the \textsc{\textit{analysis of software data}} (R142) included \textit{rapidly assessing large codebases"} (R6).



In the \textbf{writing and reporting} phase, respondents also envisioned opportunities on \textsc{formatting} (R40), \textsc{help with writing} (R57, R435, R453), and more specific support such as \textsc{feedback on one’s own writing}. \textit{Early feedback"} (R85) was described as especially beneficial for students, helping to \textit{speed up feedback cycles before manuscripts reach a supervisor"} (R204). At the same time, respondents cautioned that using GenAI as a writing assistant should not replace students’ own authorship, as this \textit{``could lead supervisors to read superficial and uninteresting prosa"} (R204).


The opportunity to support \textbf{peer review} was described as helping \textit{``a fairer allocation of [..] tasks based on reviewers' prior publications"} (R17).


When considering \textbf{GenAI as the research topic}, respondents envisioned a broader research agenda extending beyond current application-oriented studies toward systemic, methodological, and disciplinary transformation. Opportunities included understanding how SE practitioners can \textit{``get the best from it"} (R47). From the complementary perspective of \textsc{SE for GenAI}, respondents emphasized that \textit{``GenAI [is] a new kind of software. Its problems need SE methods"} (R290).

Viewed through McLuhan’s tetrad~\cite{mcluhan1977laws,mcluhan1977laws}, GenAI was further seen as \emph{retrieving} prior lines of inquiry by \textit{``compar[ing] and contrast[ing] how certain types of studies will change with the GenAI"} (R277) and reviving dormant areas of investigation, where \textit{``many topics can be revived with GenAI"} (R250). At the same time, respondents anticipated both \emph{enhancement} and \emph{reversal} in the research landscape, where GenAI may \textit{``bring new research problems for SE"} (R20) and \textit{``make others obsolete"} (R122), while also \textit{``provid[ing] new powerful tools for developers to break boundaries they couldn't break with traditional technologies"} (R122).


As a \textbf{cross-cutting} dimension, respondents 
envisioned \textsc{summarisation of text} (R52), which, while visible during idea generation, was here framed as supporting multiple research activities throughout the lifecycle.

Respondents positioned \textsc{GenAI as a research partner}, variously described as \textit{``GenAI is like a collaborator"} (R310), a \textit{24/7 research partner"} (R3), or a \textit{``thought partner"} (R58). Related opportunities for \textsc{error checking} included helping researchers \textit{``make fewer mistakes when  using] LLMs to check [the] work"} (R308), \textit{``identifying potential problems in the design of the methodology"} (R289), and \textit{``highlight[ing] areas that are overseen by the authors"} (R215).

Respondents also described expanded \textsc{access to information}, characterized as \textit{``like when Google Scholar appeared, but x10"} (R308), alongside opportunities for \textsc{``making SE and SE research more accessible}. Accessibility and inclusion were detailed as \textit{``lowering the entry barrier to SE research for non-native English speakers and those without systematic knowledge of the field"} (R187) and broader \textit{``inclusion [..] people in other disciplines"} (R332). Finally, some respondents framed GenAI’s influence as a \textsc{paradigm shift} in SE research (R300).

Respondents expressed \textbf{cynical or critical perspectives} regarding opportunities for GenAI in SE research, emphasizing ethical, epistemic, and quality-related risks.

\textsc{Ethical concerns} were mentioned, with the reflection of having the \textit{``opportunity to test the limits of the community"} (R165). Related apprehensions involved \textsc{homogeneity} and potential loss of creativity, including the need for \textit{``carefully monitoring effect on creativity"} (R85) and concerns about \textit{``better writing (but less individual [better writing])"} (R19). Respondents also highlighted risks that GenAI \textsc{introduces bias}, noting that automated feedback must be used cautiously to avoid \textit{``biased analysis [\ldots] and [effects on] creativity"} (R85), while summaries may \textit{``suffer from fixation effects"} (R19).

More broadly, several participants argued for \textsc{limited utility}, particularly for tasks requiring critical thinking. These included rejecting AI-written papers because supervisors may read \textit{``superficial and uninteresting prosa"} (R204) and doubting GenAI’s ability to generate \textit{``genuine novel ideas"} (R2). Others questioned technical progress and trustworthiness, citing limits in \textit{``reasoning and critical thinking tasks"} (R165) and uncertainty about reliability in more important tasks (R181). Some framed current benefits as restricted to \textit{``routine [\ldots] but not ‘intellectually exciting’ aspects of research"} (R56) or reported \textit{``not many opportunities"} at present (R195), including skepticism toward research-question generation (R202). Respondents also perceived \textsc{no or limited opportunities}, stating that \textit{``the problems and dangers outweigh the opportunities"} (R286), describing GenAI as \textit{``a hype bubble [\ldots] oversold"} (R155), or asserting it as \textit{``the opposite of research"} (R77). These views sometimes extended to a broader \textsc{pessimistic outlook}, framing GenAI as \textit{``a debasement of our art"} (R272), even \textit{``the down fall"} (R59), or criticizing perceived citation-driven trends and overreliance on AI-mediated research practices (R166). Others warned of long-term societal harm (R182).
In final thoughts, this pessimism extended to concerns about community identity, with one researcher lamenting that the SE research field has \textit{``become a second-class AI community''} (R166).

\textsc{Quality concerns} emphasized more risks (than opportunities) of \textit{``incorrect and shallow research"} (R74), outputs that are \textit{``more bug-prone"} (R19), and potential loss of depth among early-career researchers (R165). One respondent further cautioned that excessive reliance on GenAI may undermine trust in the research record and should be avoided in core empirical activities (R277).
Final thoughts reinforced these quality concerns, with one participant stating that \textit{``unless hallucinations are solved, GenAI is a non-starter in these contexts''} (R268), 
framing reliability not as a temporary limitation but as a blocking condition.

\subsection{How do SE researchers trust its use and what risks do they perceive from using GenAI in research, and
how do they mitigate those risks? (RQ4)}
\label{sec:rq4}

In this research question, we investigate how software engineering researchers trust generative AI in the research process, the risks they perceive in using GenAI for SE research, and the strategies they propose to mitigate these risks.

\subsubsection{How do SE researchers trust GenAI for research? (RQ4.1)}
\label{sec:rq4.1}

Researchers who \textit{use GenAI for research} reported moderate levels of trust in GenAI tools. Considering Top-2-Box scores (Agree and Completely Agree), 34\% indicated confidence in GenAI tools, 12\% perceived them as reliable for research, 9\% felt safe relying on them for research tasks, and 15\% reported liking their use for research decision-making.





\begin{figure}

        \includegraphics[width=\textwidth]{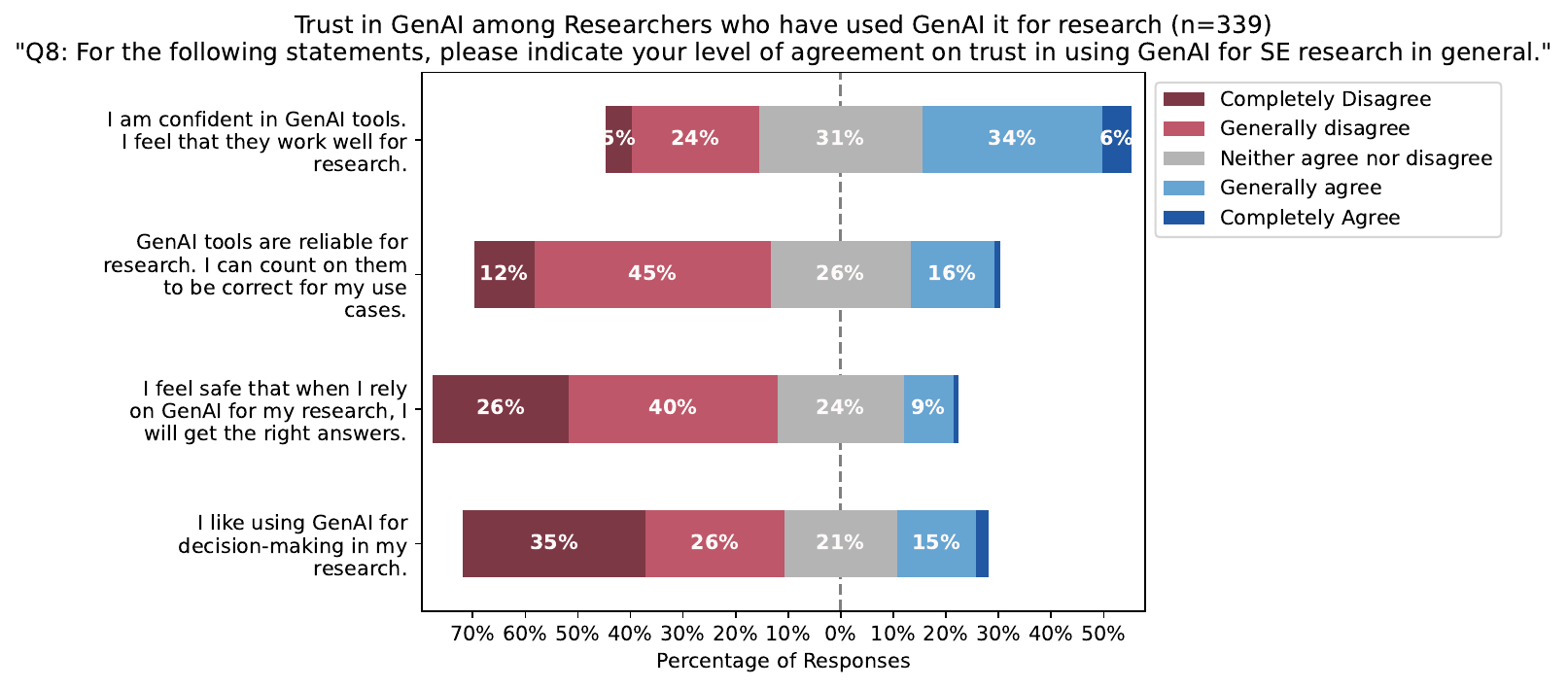}


\vspace{-4mm}
    \caption{The Different Dimensions of Trust in GenAI for SE Research (based on \cite{choudhuri2025guides})}
    \label{fig:Q8_trust_by_group}
\end{figure}

\begin{figure}
        \includegraphics[width=\textwidth]{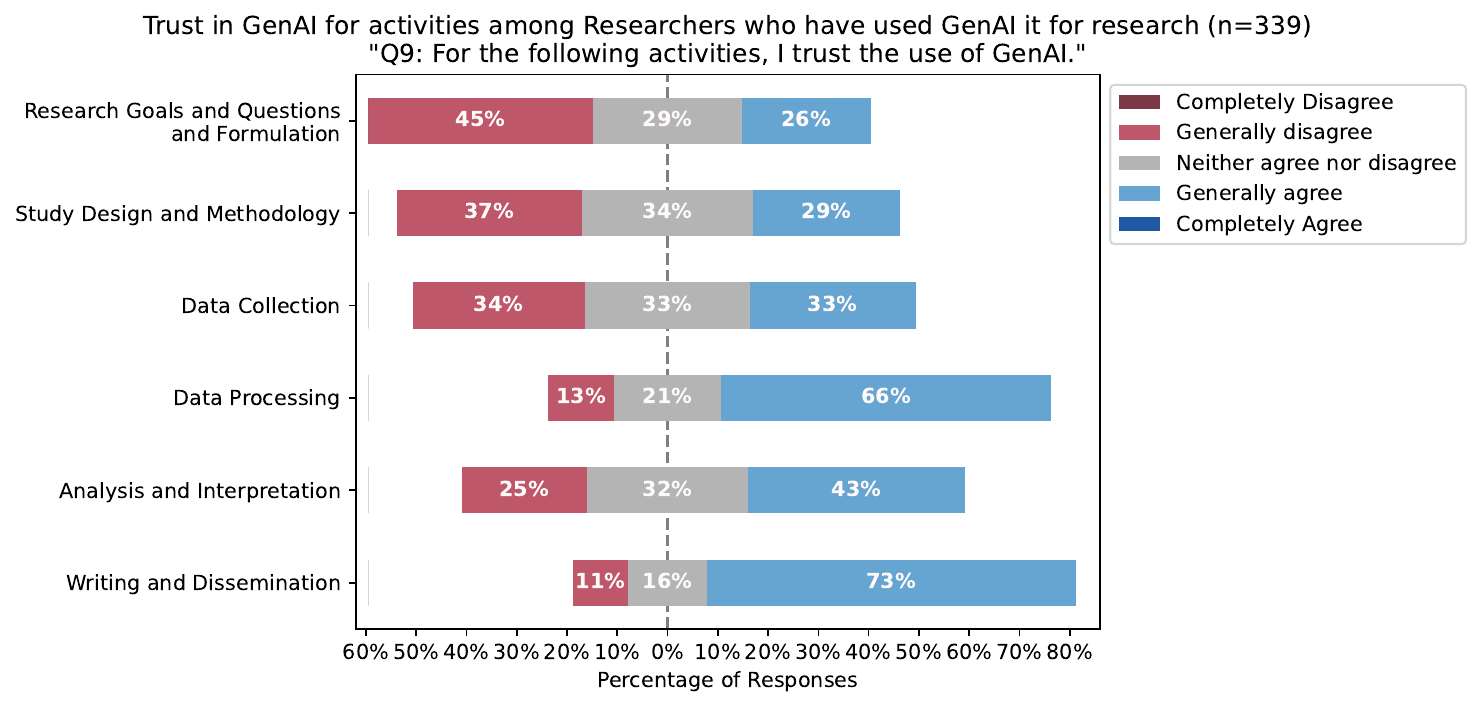}
    \vspace{-6mm}
    \caption{Trust in GenAI per Stage in the SE Research Pipeline (based on \cite{trinkenreich2025get})}
    \label{fig:Q9_trust_by_group}
\end{figure}

\textbf{Trust in GenAI across research activities:} As shown in Figure \ref{fig:Q9_trust_by_group}, trust in GenAI varies across different stages of the research process and across user groups.

Researchers who \textit{use GenAI for research} reported higher levels of trust in later-stage research activities. Trust was highest for \textit{Writing and Dissemination} (77\%), followed by \textit{Analysis and Interpretation} (41\%), \textit{Data Collection} (31\%), \textit{Study Design and Methodology} (25\%), and \textit{Research Goals and Question Formulation} (23\%).





\subsubsection{How do SE researchers perceive the risks of using GenAI for research? (RQ4.2)}
\label{sec:rq4.2}
\par\noindent 
This research question examines the perceived risks of using GenAI in SE research.

\begin{figure}
        \includegraphics[width=\textwidth]{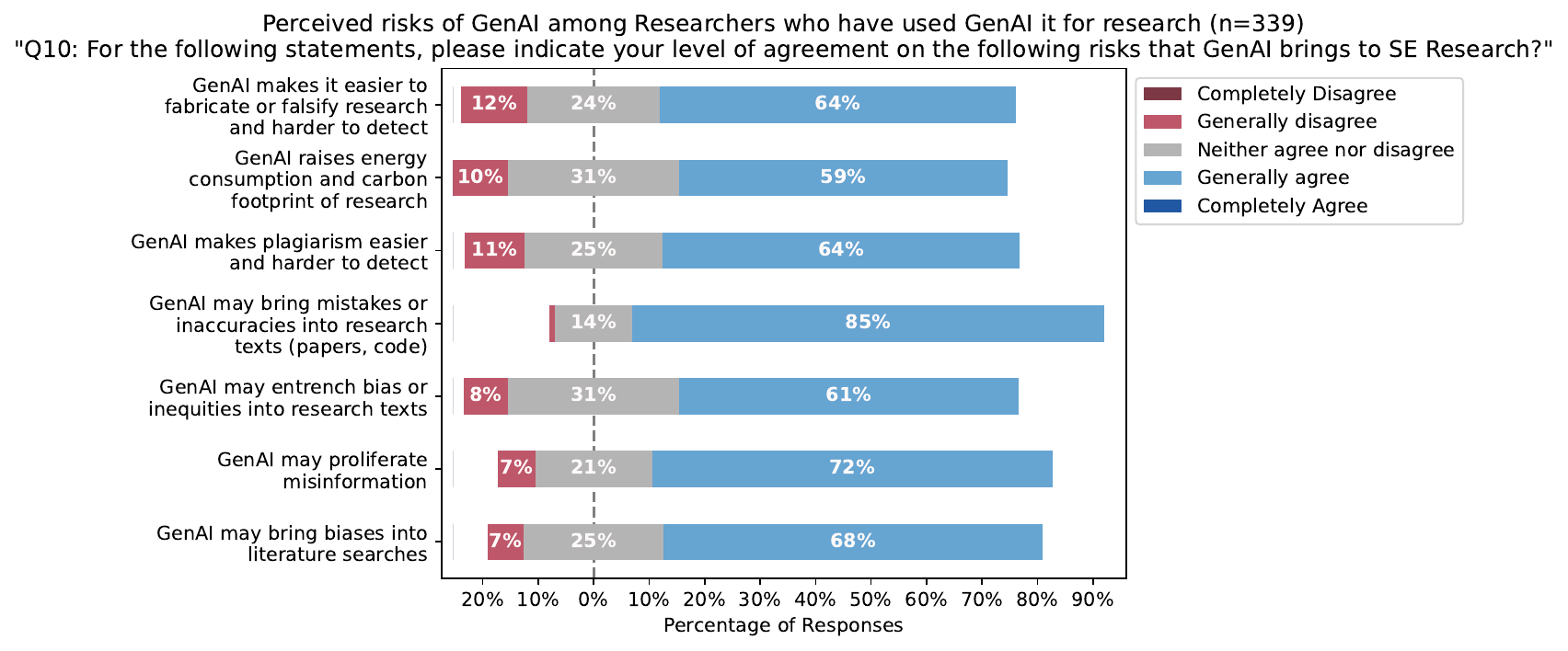}
    \vspace{-6mm}
    \caption{Perceived risks of using GenAI for SE research}
    \label{fig:Q10_risks}
\end{figure}

Figure~\ref{fig:Q10_risks} presents the perceived risks associated with the use of GenAI in SE research.


Considering Top-2-Box scores (Agree and Completely Agree), researchers who use GenAI for research 
are mostly concerned about the possibility that GenAI may introduce \textsc{mistakes or inaccuracies into research texts}, though the majority agreed with all of the potential risks. 

\subsubsection{How can SE researchers mitigate the risks without missing the opportunities GenAI offers? (RQ4.3)}
\label{sec:rq4.3}

\par
\begin{figure*}[t]
\centering
\includegraphics[width=0.8\textwidth]{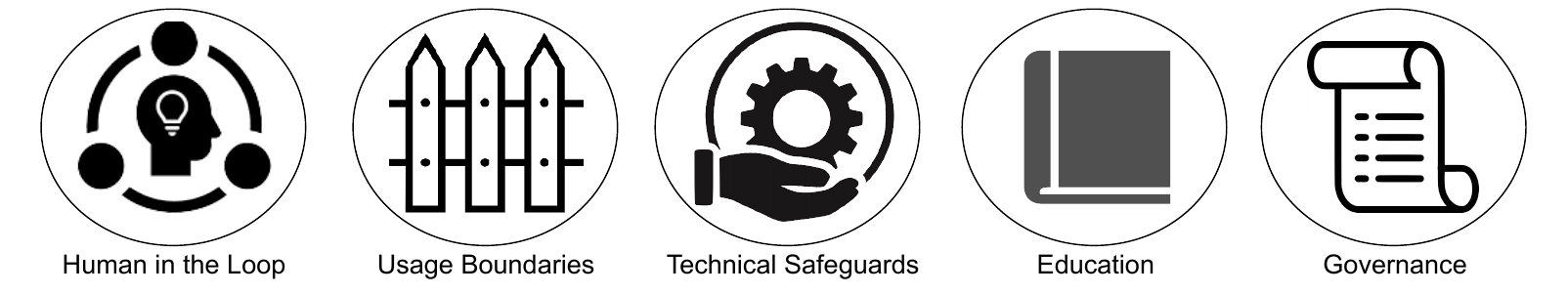}
\vspace{-3mm}
\caption{The mitigation strategies reported by SE researchers who participated in our study.}
\label{fig:mitigations}
\end{figure*}

\noindent 
This research question examines the actions participants identified to mitigate the perceived risks of GenAI in SE research. Our analysis revealed five categories of mitigation strategies, as illustrated in Fig.~\ref{fig:mitigations}.

Table \ref{tab:codes_mitigations} presents the number of participants whose responses fit in each category. In the following, we present more details about our findings, organized by category of mitigation strategy. Because participants’ responses could reflect multiple mitigation strategies, individual responses were sometimes coded into more than one category.

\begin{table*}[b]
\centering
\small
\caption{Representative examples of answers to the risk mitigation open question, number and percentage of respondents whose answer was coded for each category.}
\label{tab:codes_mitigations}
\begin{tabular}{c|l|r|r}
\hline
\toprule
 \begin{tabular}[c]{@{}c@{}}\textbf{Mitigation} \\\textbf{Strategy}\end{tabular} & \textbf{Representative examples} & \textbf{\#}  & \textbf{\% (n=138)}  \\
 \hline

\toprule
\begin{tabular}[c]{@{}c@{}}Human in \\the Loop \end{tabular}& \begin{tabular}[l]{@{}l@{}}\textit{``AI does not substitute the human, who should inspect everything"} (R62)\\ \textit{``The researcher has the final word. They have to review what the GenAI generates truly.} \\\textit{We cannot blindly trust it."} (R135)\end{tabular} & 90 & 65.2\% \\ \hline

\begin{tabular}[c]{@{}c@{}}Usage \\Boundaries \end{tabular}& \begin{tabular}[l]{@{}l@{}}\textit{``Researchers should use GenAI as a tool to help with specific, well-defined tasks [..]}\\\textit{They should not attempt to "offload at the edge" of knowledge tasks to a text completion} \\\textit{engine."} (R42)\end{tabular} & 11 & 8.0\% \\ \hline

\begin{tabular}[c]{@{}c@{}}Technical \\Safeguards \end{tabular}& \begin{tabular}[l]{@{}l@{}}\textit{``RAG-based usage,"} (R164), \textit{``agentic systems to assist in verifying}\\\textit{research,"} (R138) \textit{``use separate conversation thread and modify the}\\ \textit{question to see if the answer changes"} (R172)\end{tabular} & 13 & 9.4\% \\ \hline

Transparency & \begin{tabular}[l]{@{}l@{}}\textit{``Disclose the usage of AI in research in more detail,"} (R427), \textit{``mak[ing]}\\ \textit{it explicit which parts of the research were supported} by them (e.g., Threats to Validity \\\textit{section)"} (R198)\end{tabular} & 17 & 5.9\% \\ \hline

Education & \begin{tabular}[l]{@{}l@{}}\textit{``Tell and demonstrate [researchers] what happens if [they] use GenAI incorrectly,}\\ (R107) \textit{``Promote ethical behaviour. Encourage the detection}\\\textit{(and public report) of published materials corresponding to illustrative cases of}\\\textit{misinformation, falsification, and plagiarism."} (R159)\end{tabular} & 14 & 10.1\% \\ \hline

\begin{tabular}[c]{@{}c@{}}Governance \\Controls\end{tabular} & \begin{tabular}[l]{@{}l@{}}\textit{``Open science and replication requirements,} (R74) \textit{``design new}\\\textit{trustworthy assurance and evaluation methodolygy"} (R15)\end{tabular} & 4 & 2.9\% \\ \hline
\bottomrule
\end{tabular}

\begin{center}
\parbox{0.9\textwidth}{%
\footnotesize
The total per mitigation strategy is not the sum of the respondents since participants often provided an answer that was categorized into \textbf{more than one strategy}.
}
\end{center}

\vspace{-0.2cm}

\end{table*}


\textbf{Human in the Loop} is an interaction paradigm in which AI-generated outputs are treated as suggestions that remain subject to human judgment, validation, and accountability rather than being accepted autonomously. Respondents emphasized that \textit{``GenAI should not be used to replace researchers, but can only help them"} (R47), highlighting that responsibility and agency must remain with humans.
This view was echoed in final thoughts, where one respondent cautioned that \textit{``entirely GenAI generated feedback to papers and draft is wrong''} and that \textit{``GenAI should be used to complement what humans do''} (R279). 

Respondents stressed the need to \textsc{``double check GenAI results"}, emphasizing that outputs should be verified rather than used blindly. Across responses, double-checking was consistently associated with manual inspection, cross-verification with external sources, and the expectation that \textit{``the human must make the final decision and take the responsibility"} (R326). Some articulated concrete verification practices, such as comparing with \textit{``traditional searching solutions"} (R455) and ensuring that \textit{``citations [..] point to clear parts of the cited manuscripts"} (R22). Double-checking include criticizing GenAI results (R449). Respondents argued for being \textit{``quite critical, analytical, and aware of whatever GenAI might generate"} (R435), warning that even common academic-writing support can introduce substantive inaccuracies. For instance, GenAI-generated text for an introduction or abstract may, \textit{``in an attempt to create something novel,"} \textit{``hallucinate and include aspects that aren't actually in the paper"} (R435). Similarly, when used for related-work analysis, it can be misleading because \textit{``it even generates its own information that isn't associated with the referenced papers"} (R435).
The final thoughts reinforced this verification norm, with one researcher framing it as a professional duty \textit{``[\ldots] to double check if the suggestions or pointers provided by GenAI are actually factually true''} (R343). 
Others provided verification workflow examples, such as routinely confirming \textit{``if the rewritten text is still semantically equivalent''} before accepting GenAI-rewritten prose (R34). 


\textit{\textsc{``Maintain human oversight"}}, \textit{``especially in tasks involving interpretation or critical decision-making"} (R3), was highlighted as a process in which \textit{``all the content [is] properly verified before usage"} (R76), in which researchers may \textit{``ask genAI and then decide"} (R75), but be supported by \textit{``manual analysis"} (R16) of the output. In this framing, GenAI is explicitly treated \textit{``as a tool, mak[ing] sure you look at what it's saying and use your best judgement"} (R7). Respondents emphasized retaining human ownership of outcomes, stating that humans should have \textit{``the first and the last word"} and that GenAI should be used only \textit{``as a auxiliar tool"} (R247). In order to maintain human oversight, respondents suggested that GenAI \textit{``must be used by people who can already perform those tasks effectively"} (R184).
The final thoughts corroborated this stance. One respondent stated that \textit{``GenAI is like any other tool---the responsibilities stay with humans using it''} (R343), 
while another  commented that \textit{``papers and reviews are the products of humans, and humans are responsible for accuracy''} (R336).

\textsc{Review results more carefully} emphasizes deeper, more skeptical evaluation of GenAI-assisted outputs by both researchers and reviewers. Respondents stressed the need to \textit{``remain cautious about the genAI results"} (R17) and argued that reviewers would need to \textit{``take more time on reviewing manuscripts manually and in more detail,"} described as the only way to address risks such as falsified results, plagiarism, and inaccuracies (R19). This stance also involves preparation before tool use, with one respondent noting the importance to \textsc{run manual analysis before using GenAI} by \textit{``preparing content or having an idea about the topic [..] and after reviewing the results carefully"} (R61). Although acknowledged as difficult to scale, respondents associated careful review with heightened rigor, emphasizing that \textit{``the bar on research rigor must be raised"} (R68).
One researcher's final reflection  underlined this expectation, stating that \textit{``the use of AI tools is not an excuse for inadequate review''} (R336).


\textbf{Usage Boundaries} refer to strategies that constrain when, how, and for which purposes GenAI should be used. Rather than treating GenAI as a general-purpose assistant, respondents emphasized the need for concrete boundaries and data protection. Restrict Task-Types was mentioned as a boundary to use of GenAI on tasks that do not require deep understanding of methodological judgment. For example, in literature reviews, while \textit{``summarizing existing literature is a good thing,"} researchers \textit{``shouldn't always rely on GenAI for selecting papers,"} emphasizing that \textit{``true personal understanding can happen by actually reading the paper and not a summary"} (R2). Respondents suggested using GenAI to \textit{``help with English grammar)"} (R42), \textit{``brainstorm paper titles or keynote titles,"} (R52) and specific parts for the paper, as a \textit{``draft part of an introduction,"} emphasizing that the generated text is treated as provisional scaffolding rather than the final prose (R167). In this approach, GenAI-generated content is explicitly framed as temporary that would \textit{``ultimately get replaced before submission"} (R167).


\textbf{Technical Safeguards} were described as tools and automated checks that support the evaluation of GenAI-produced results. At the system level, respondents advocated for \textit{``mak[ing] AI systems resemble traditional systems"}. At the tool-selection level, respondents emphasized choosing models that better support reliability. For example, one mentioned \textit{selecting LLM that creates trustable summary (e.g. NotebookLM\footnote{\url{https://notebooklm.google.com/}})"} (R141). Beyond generation, respondents also proposed augmenting evaluation workflows. Several respondents suggested using an automatic tool to identify \textit{``misinformation or incorrect claims''} (R20). Respondents also stressed context-specific safeguards, noting the need to \textit{remain careful with using GenAI for any actual data processing [when] work[ing] with personal data"} (R192).


\textbf{Transparency} captures mitigation strategies that emphasize making GenAI use, decision processes, and research artifacts visible and inspectable to researchers, reviewers, and the broader community. Respondents stressed the importance of \textit{``clearly documenting [the GenAI] role in writing, analysis, or ideation"} (R3), including making explicit which parts of the research were supported by GenAI (R198). In addition, respondents highlighted the need to \textit{``make the decision making more transparent"} (R145) and argued that \textit{``in order to trust GenAI tools, the underlying models should be transparent and customizable"} (R151). From this perspective, transparency also involves making system behavior interpretable, such that tools \textit{``reflect the context behind the information"} (R152).


\textbf{Education} suggestions emphasized the role of cultural norms and community values in shaping responsible GenAI use. Several participants explicitly called for education and awareness, stating \textit{``Education, education, education!"} and urged researchers to be shown \textit{``what happens if you use GenAI incorrectly"} (R107). This also involves broader awareness-building, including \textit{``mak[ing] the researchers and teachers and students aware of [GenAI] risks"} (R340) and of \textit{``the perils of GenAI"} (R166). Finally, respondents linked education to structural changes, developing \textit{``techniques that helps researchers to use these tools correctly and with ethics perspective"} (R411).


\textbf{Governance Controls} are strategies about institutional, procedural, and regulatory mechanisms to control how GenAI is used in research. A strategy that stays between training and governance to know \textit{``what is or is not ok to use AI for"} (R83). Others extended this responsibility to institutions, arguing that \textit{``publishing venues and funding agencies need proper guidelines and quality assurance measures"} to prevent and detect GenAI-related issues such as plagiarism (R158).

\subsection{What regulations and policies do researchers feel should be applied to the use of GenAI in their research and in peer review? (RQ5)}
\label{sec:rq5}

In this research question, we examine the regulatory and policy needs identified by SE researchers regarding the use of GenAI in research activities.



In the survey we asked an open-ended question: ``Should GenAI use be regulated in Software Engineering research, assuming that it is possible?''.  We present the overall stance of respondents, followed by three themes about how regulation that emerged from the responses. 

\paragraph{Overall Stance Towards Regulating GenAI Use in SE Research}


From the 119 participants who answered this question, the majority (101) suggested that ``Yes'' GenAI should be regulated, 8 were more on the fence, and 9 tended towards ``No'', it should not be regulated at all. 

While the majority agreed that some regulation is needed, they also qualified their response with details on which kinds of research tasks regulation is needed (e.g., reviewing, data analysis), or that new researchers may be in need of more regulation (or guidelines).  The reason given was not just about the integrity of the research, but also because, automating research reduces the learning opportunities especially for new researchers: \textit{``It should not be used in Systematic Literature Reviews. Why? Maybe it's a personal bias, but I have learned in the last, say 5 years, how to perform an SLR and just few days ago I so in a course how easy is to do this with the help of AI. BUT, I think GenAI steals the opportunity of the researcher/learner to learn! The same in teaching: if somebody else (even GenAI or another human, friend, parent) does the task instead of you, that person will learn and not you!''} (R340)

For those who were against using GenAI at all, they had even stronger opinions, for example: 
\textit{``I believe if a researcher is using GenAI to think for them, they might as well remove their PhD title from their CV.''} (R166)
While others felt that at the minimum its use should be regulated to be transparent, but even then boundaries should be established in how it is used: \textit{``Clear statement when GenAI is used as part of the research (as an assistant to help in a specific SE task or when it is used as a tool to deliver novelty (but in this case, it should not be allowed).''}(R340)

For the few that were strongly against any kind of regulation, they brought in parallels to the use of Google and the internet in our earlier research: 
\textit{``The use of GENAI as a coding and writing assistant is prevalent and does not violate any code of ethics to my knowledge. I DON'T THINK ANY REGULATION IS NEEDED. If Google era research was never regulated then GENAI era research should be no different. GENAI is only as good as a user's prompt. Since prompts are a user's personal intellectual property so the resulting output which is generated as a result of the prompt also belongs to the prompt writer. Perhaps I do not have any misuse cases in my mind right now otherwise I would have been better able to comment on this question.''} (R171)

One respondent even considered us asking this question was conceptually flawed: 
\textit{``Honestly, I find that a difficult premise to accept. Let's imagine that access to widely-used cloud services like ChatGPT, Gemini, and others is suddenly restricted. As long as we remain committed to the open principles of the Internet, anyone with reasonably capable hardware should still be able to train their own models and use them for research. In fact, I believe the question itself is conceptually flawed. Indeed, it assumes that we would willingly forgo our free will (assuming it exists), which runs counter to the spirit of scientific progress. Anyhow, you guys should definitely asks ChatGPT or whatever models you feel like using :)''}(R73)

\vspace{.2in}
Three themes about regulation, if applied to AI use in research, emerged from the responses.



\paragraph{Regulation requires human judgment} 

A cross cutting theme across many of the responses is that just as human input is used during GenAI use, it is also often needed as its use is regulated, but some mentioned that 
our community really needs to ``lean in'' to using GenAI but should use regulation wisely: \textit{``Yes, but I would like to see some kind of smart regulation, which enables innovation while minimizes harm. No blunt restrictions''} (R58).
Researchers' final thoughts reinforced this stance, with one respondent noting that \textit{``prohibiting these tools is just impractical''} and another called instead for \textit{``a culture of scientific integrity with respect to their use''} (R219).


\paragraph{Epistemic \& Scientific Integrity Concerns}

For those that answered it should be regulated,  many participants mentioned concerns about the rigor, reproducibility, and transparency of research: \textit{``Yes, GenAI use in Software Engineering research should be regulated to ensure ethical practices, transparency, and reproducibility, while allowing room for innovation and development''} (R142). There were further concerns about research integrity, notably misinformation risks (from biases and hallucinations) and correctness or accuracy of the research results supported by GenAI: \textit{`` I think there's a need to regulate because I've reviewed papers where it appeared to me that the author had copied text directly out of ChatGPT and it made me suspicious that there could be inaccuracies or hallucinations in the paper.  But I really don't know how you go about regulating it.  That's a super tough problem.  Seems like someone should do some research on solutions''} (R7).


\paragraph{Governance, Responsibility and Accountability}

In addition to concerns about misinformation (mentioned above), there were concerns about the environment: \textit{``...GenAI use can introduce bias, security flaws, misinformation, and environmental impacts. Regulation would encourage responsible usage and proactive risk management''}(R23). Public trust in our research was also mentioned:  \textit{``Yes, to the extent that it is regulated in other fields. I believe it is okay to use generative AI for writing code to preprocess data or implement models, for example, as long as authors (a) check the code carefully and (b) are aware that they are liable for any mistakes the model makes that they did not detect. However, I am generally against its use in other parts of SE (and most other fields') research. It is also important that as the scientific community, we maintain public trust in science, which may be harder if generative AI use were to be widespread''} (R21). 

In the final question in the survey on their final thoughts, some researchers offered concrete proposals for how such governance could be operationalized. 
Some respondents underlined the challenge that \textit{``it is not always possible to create generic guidelines, as each domain is bound by certain issues''} and suggested that \textit{``guidelines should be created based on the type of study, such as the SIGSOFT [empirical standards]''} (R197).  Others proposed a task-based decision matrix specifying acceptable GenAI use per research activity, for example \textit{``forbidden for reviewing, accepted for brainstorming, [allowed] for data labeling only when fulfilling a set of very specific requirements''} (R19). To sustain such efforts, respondents called for dedicated sessions at major venues such as ICSE, FSE, and ASE to discuss community rules, emphasizing that \textit{``this needs to be done continuously''} through \textit{``dedicated boards [\ldots] at all conferences and journals''} (R19).

\section{Discussion}
\label{sec:discussion}

\subsection{Emergent Tensions Across Findings}

\subsubsection{Productivity Tensions}
Productivity gains were one of the most commonly mentioned opportunities of using GenAI. However, the potential productivity benefits were often tempered by a set of emerging tensions. 

\paragraph{Productivity-Effort Tension}
Participants emphasized that using GenAI well is far from trivial, as \textit{``it takes extra effort and experience for researchers/reviewers to gain confidence, rather than creating shortcuts and bad outcomes''} (R336). 
One respondent captured this tension in terms of expertise: \textit{``GenAI tools can be very useful in the hands of skilled researchers who know what they are doing and can distinguish between right and wrong information. However, in the hands of a novice, it has the potential to wreak havoc, leading to fabricated research and spreading misinformation''} (R82).Another stated that \textit{``using GenAI sensibly requires a lot of effort that many researchers are not willing or capable of investing''} (R158). These responses point to a paradox. The primary appeal of GenAI is efficiency, yet responsible use demands significant effort, expertise, and critical engagement. However, if the effort barrier is not addressed through training and community norms, the productivity promise of GenAI risks being realized at the expense of research quality.

\paragraph{Productivity-Quality Tension}

Quality concerns were also explicitly raised. One participant felt that GenAI output was \textit{``potentially biased, incomplete, or suffering from fixation effects''} (R19). Another participant stressed that diminishing research quality affects the entire SE research community: 
\textit{``As community, we build upon previous research, we extend them, we corroborate them, we contradict them, and that's how we have moved forward as a community. We always trusted previous research. But now I am at a juncture, where I cannot decide whether I trust the research anymore.'' }(R277)

These concerns illustrate a second tension: while GenAI may increase the volume or speed of research production, it may simultaneously undermine confidence in the integrity of the research ecosystem that relies on shared trust and cumulative progress.

\paragraph{Productivity-Impact Tension} We also found evidence that researchers who use GenAI felt pressure to do so both to stay relevant and to secure funding (Sec.~\ref{sec:rq1.2}. This pressure may increase the amount of research being produced, but can also potentially skew agendas toward what is ``AI‑adjacent,'' raising questions about long‑term scientific impact vs. short‑term output. One participant described the potential for more shallow research:  \textit{``I think it will result in incorrect and shallow research in many cases.''} (R74) This raises a third tension: long term research impact may be at risk when GenAI is used to amplify productivity alone.

\smallskip

These tensions reinforce the calls for human-in-the-loop (Section~\ref{sec:rq4.3}) and governance (Section~\ref{sec:rq5}) mitigation strategies  reported earlier in our findings.

\subsubsection{Reliance-Control Tension}

A related but distinct tension concerns \emph{who} is behind the wheel steering the research process, and whether academics risk becoming passive consumers of GenAI outputs rather than active producers.
Several respondents warned against giving over too much control to GenAI. One cautioned that researchers should \textit{``be careful of relying too heavily on it,''} arguing that \textit{``there's a lot of value in thinking deeply about the research and forming a strong understanding of your data, which could easily get lost by giving GenAI the wheel''} (R331). Another observed that \textit{``the excessive use of GenAI by students and researchers shifts the actual research to the AI''} (R77). A concrete example of this loss of control was offered by one respondent who reported that \textit{``authors of the paper that I recently reviewed apologized for paragraphs that were `AI-massaged,' so the content was incorrect,''} concluding that \textit{``GenAI might make us lazier and less responsible for the things we write''}
(R161). These responses reveal a tension between the convenience of delegation and the oversight that responsible use of GenAI demands. Without proper mitigation, over-reliance can occur. This tension also reinforces the human-in-the-loop mitigation strategy (Section~\ref{sec:rq4.3}).

\subsubsection{Democratization-Inequity Tension}
While participants noted that GenAI has the potential to make SE research more accessible, a lack of computing resources was also cited as a concern by many participants. These sentiments lead to another emerging tension. The cost of using the latest models on large scale SE data can become astronomical. While GenAI may make some information more accessible, it can also be leveraged best by those with large research budgets. This can lead to even greater inequities if reviewers demand the use of the latest models and more data in experiments. This also reinforces the need for governance (Section~\ref{sec:rq5}) to ensure reviewers do not amplify inequities.

\subsubsection{Rapid Change-Stable Guidance Tension}
Another tension that emerges is related to the need for stable guidance and regulations, balanced with the rapid change of the GenAI tooling landscape. Respondents discuss the need for guidance, rules, and regulation for GenAI use, a theme developed in the risk mitigations (Section~\ref{sec:rq4.3}) and in the calls for regulation and policy (Section~\ref{sec:rq5}). Each of the tensions mentioned above ultimately ties back to this need for shared expectations about how GenAI should be used responsibly in SE research. Yet, the GenAI landscape is rapidly changing. As a result, researchers are navigating a moving target: they desire durable community guidance, but any rules that are too prescriptive risk becoming outdated almost immediately. This creates a tension between the stability needed to maintain research quality and integrity, and the flexibility required to respond to continual technological shifts. The SE community must pursue guidance that is principled rather than tool‑specific, focusing on research values—such as transparency, verification, and accountability—so that expectations remain relevant even as GenAI technologies continue to evolve.

\subsection{The Competence Pipeline at Risk} 

While our participants highlighted many benefits and opportunities of using GenAI in SE research, as final thoughts many also cautioned that human expertise is needed to evaluate the outputs and leverage these benefits. For example, when describing GenAI opportunities, R63 said \textit{``It's a great helper if and only if you already have a fundamental understanding.''}
One participant (R165) said, \textit{``The major threat of using GenAI summaries is to loose depth, particularly for young researchers''}. These sentiments raise questions about the development of these fundamental research skills when students begin their research careers with GenAI.  When sharing final thoughts, nine
respondents explicitly pointed to the potential erosion of the research training pipeline, a concern that cuts across the risk findings (Section~\ref{sec:rq4.2}) and the education-related mitigation strategies (Section~\ref{sec:rq4.3}).

Five respondents raised concerns about the impact on education and training. One expressed worry that \textit{``GenAI will enable researchers to conduct a significant portion of the research by themselves,''} with the consequence that \textit{``researchers will require less low-level support from undergraduates and graduate students because it will take so much more time to teach them how to do and inspect the output [\ldots] than doing [it] themselves''} (R62).
This dynamic suggests that faculty may prioritize short‑term productivity over mentorship, reducing students’ opportunities to develop research skills through hands‑on practice.

Related to this, four respondents highlighted the risk of skill degradation. One noted that \textit{``the use of GenAI moves researchers a step further away from the research. You don't have to get your hands dirty in the data because a model can just summarize it''} (R7).
Another captured the tension between improved surface quality and diminished understanding: \textit{``maybe the English quality is becoming better and better, but we don't know what we're writing anymore''} (R161).

At the same time, respondents recognized that the solution lies not in avoiding GenAI but in rethinking how the next generation of researchers is trained. R23 argued for \textit{``educating the next generation of researchers, not just on how to use GenAI tools effectively, but also how to critically evaluate, audit, and even design them responsibly.''}

These findings suggest a cumulative risk: if researchers lack the training to use GenAI critically, they may be more likely to produce lower-quality outputs, which could erode trust in AI-assisted research. 
Addressing this risk requires integrating GenAI literacy into research training, as discussed in Section~\ref{sec:rq4.3}.

\subsection{Implications for Practice and Research Opportunities}
The tensions described above have several implications for software engineering researchers. First, the Productivity-Effort and Productivity-Quality tensions highlight that researchers should not expect efficiency gains without friction. Researchers must budget time for skill acquisition and embed QA processes when adopting GenAI tools as part of their research practice. Human oversight should be baked into all tasks where GenAI is used to ensure reliable and trustworthy research outputs.
In addition, the community would benefit from shared experimentation.  More studies should be conducted that evaluate GenAI's strengths and limitations for various research tasks. Such studies should not only report accuracy or speed benefits, but reflect on how human oversight is embedded, the friction points that were encountered, and other learnings that can help to collectively reduce the barrier for responsible adoption and develop guidelines for methodological design. 

The competence pipeline risk described above risks degrading researcher skills if GenAI is overused and impacts research skill formation. This can further amplify the productivity-quality tension if skills erode to the point where competent human oversight is at risk. Research supervisors and graduate programs must explicitly design opportunities for students to practice skills such as critical thinking, data collection and analysis, and writing without GenAI assistance. Developing strong foundational skills will help students leverage the productivity gains through GenAI use responsibly. 

Our results illustrate that researchers are using GenAI across all stages of the research pipeline, from idea generation to writing and reporting. This further amplifies the call for the need for methodological guidance for responsible GenAI use across the research pipeline. The community needs shared norms, templates, and standards on when and how GenAI can be used, how it should be validated, and what should be reported to preserve transparency and replicability. 

Notably, research design and data collection were the least commonly reported phases for GenAI use. These phases often require judgement and reasoning. Beyond better guidelines, we envision tooling that could be helping in aiding researchers to leverage GenAI in these use cases. For research design, specialised GenAI tools with knowledge of the ACM SIGSOFT empirical standards could be created to help guide researchers through the research design process. Rather than replacing human reasoning, the tools could be used as research partners helping to drive methodological structure in the research design phase. Being grounded in the existing empirical standards can also help to ensure the research design meets the expectations for soundness and rigor. 

Similarly, tools could be envisioned to support data collection, though safeguards will be needed. The type of tool will depend on the source of the data being collected. For example, mining software repository studies, GenAI can be used to generate data collection scripts and tools could be created to verify the accuracy of collected data. Such verification tools might automatically detect inconsistencies, flag suspicious patterns that suggest faulty scraping or API failures, or compare GenAI‑generated extracted data against known repository data or structures. These kinds of guardrails would help ensure that GenAI augments data collection without introducing silent errors or biases. For data collection involving human subjects, such as interviews or questionnaires, GenAI could be used to pilot collection instruments to ensure misleading or biased questions are identified. However, GenAI cannot replace genuine perspectives of human participants, and caution is needed to ensure that the use of GenAI, even in piloting, does not introduce bias into the study design.

\section{Threats to Validity}

We discuss the main threats to the validity of our study, organized along the categories proposed by Wohlin et al.~\cite{wohlin2012experimentation}.

\paragraph{Construct validity}
A potential threat concerns whether the survey instrument adequately captures the construct of GenAI use in research. One respondent noted in the final thoughts question that the questionnaire \textit{``did not transport [the] distinction well''} between using GenAI to assist the research process (e.g., writing, brainstorming) and using it as a research tool (e.g., generating test cases or analyzing data) (R204).
We acknowledge that collapsing these distinct modes of use into a single set of questions may have reduced construct precision. Future refinements of the instrument could introduce separate question blocks for process- and tool-level assistance.

\paragraph{Internal validity} 
One threat to internal validity concerns the qualitative coding process. 
For each open-ended question, coding was performed by one author and subsequently reviewed by three other co-authors. 
Disagreements were resolved through meetings. 
While this process mitigates individual bias, having a single initial coder may have influenced the framing of each codebook.
To support transparency and replicability, the codebooks and coded data are included in the replication package.

\paragraph{External validity}
Several respondents questioned the SE-specificity of the study in the final thoughts question. One stated that \textit{``the use of GenAI in SE research would [not] be any different from using it in any other field of computer science, or even beyond''} (R182), while another argued that \textit{``many other researchers in other areas are using GenAI''} and called for studying \textit{``the wider impact of GenAI on research activities''} (R171). We acknowledge that many of our findings, particularly those related to productivity, trust, risk perception, and governance, likely reflect dynamics common across academic disciplines rather than being unique to SE. However, the purposive sampling strategy, targeting authors from leading SE venues, ensures that the findings are grounded in a well-defined research community. 
Investigating the extent to which our findings generalize to other disciplines remains a promising direction for future work.

\paragraph{Conclusion validity}
In the final thoughts question, one respondent suggested that additional questions could have probed \textit{``what new SE-related problems [are] introduced by GenAI''} (R20), pointing to potential gaps in the topic coverage of our instrument.
While the open-ended question compensated for such gaps, future studies could benefit from a broader set of questions targeting emergent concerns.

\section{Conclusion}

This paper presents the results of a large-scale survey of 457 software engineering researchers, providing an empirical characterization of how GenAI is being adopted and perceived across the SE research community.

Our findings show that GenAI adoption is already widespread, with nearly three-quarters of respondents reporting its use for research. Yet this adoption is uneven across research stages. It usage is concentrated on writing support, summarization, and coding, while research design and data collection see much lower adoption of GenAI. Data analysis falls somewhere in between these two: although reported less frequently than writing, the qualitative responses in the survey reveal it is used for annotation and exploratory analysis, typically with human oversight of interpretation. Overall, researchers tend to delegate routine and repetitive work to GenAI while retaining control over tasks that require methodological judgment. Trust follows a similar pattern, with writing and dissemination receiving by far the highest trust levels.

Our findings also uncover a set of tensions regarding the use of GenAI in SE research. Productivity gains, a recurring theme across perceived benefits and opportunities, are contrasted with concerns about the effort required for responsible use, the risk of quality degradation, and the possibility that institutional pressures may incentivize volume over depth. The competence pipeline is a related concern: if early-career researchers come to rely on GenAI before developing foundational skills in critical thinking, data analysis, and academic writing, the community risks eroding the very expertise on which responsible GenAI use depends.

Regarding governance, almost all respondents agreed that some regulation is needed, though they stressed it should focus on principles rather than specific models. Peer review emerged as a particularly divisive topic, with views ranging from outright prohibition to conditional acceptance with safeguards. These positions reflect the difficulty of establishing norms, also because GenAI technology is still evolving rapidly.

Finally, we contribute taxonomies of GenAI use cases, opportunities, risks, and mitigation strategies grounded in SE researchers' own practices and perspectives. Together with the quantitative characterization of adoption, trust, and perceived impact, these findings provide an empirical baseline against which future shifts in practices, perceptions, and policies can be measured.

Looking ahead, we see three priorities. First, the SE community would benefit from shared, evolving guidelines for GenAI use across the research pipeline, grounded in transparency, verification, and accountability. Second, graduate training programs should be designed to ensure that students develop strong research skills alongside GenAI literacy, not as a substitute for it. Third, longitudinal studies are needed to track how the tensions identified in this paper play out as GenAI capabilities evolve and community norms are developed.

\section{Acknowledgements}
We would like to thank all of our participants who dedicated their time to answering the questionnaire.

\bibliographystyle{ACM-Reference-Format}
\bibliography{references}

\end{document}